\documentclass[useAMS,usenatbib,a4]{mn2e}

\usepackage{float}
\usepackage{graphicx}
\usepackage{amssymb}
\usepackage{amsmath}
\usepackage{datetime}
\usepackage[normalem]{ulem}
\usepackage{times}
\usepackage[export]{adjustbox}

\usepackage{color}
\usepackage{soul}
\definecolor{stan}{rgb}{0,0,1}

\definecolor{dhc}{rgb}{0,0,1}

\definecolor{jon}{rgb}{0,0,1}

\def\<<{{\ll}}
\def\>>{{\gg}}

\def\spose#1{\hbox to 0pt{#1\hss}}
\def\ltwig{\mathrel{\spose{\lower 3pt\hbox{$\mathchar"218$}}
     R_{\rm A}ise 2.0pt\hbox{$\mathchar"13C$}}}
\def\gtwig{\mathrel{\spose{\lower 3pt\hbox{$\mathchar"218$}}
     R_{\rm A}ise 2.0pt\hbox{$\mathchar"13E$}}}
\def\+/-{{\pm}}
\def\=={{\equiv}}
\def\mubar{{\bar \mu}}
\def\mustar{\mu_{\ast}}

\def\Rstar{R_{\ast}}
\def\Bstar{B_{\ast}}
\def\Mstar{M_{\ast}}

\def\Mdot{\dot M}
\def\mdot{\dot m}

\def\solar{\odot}
\def\Msun{M_{\solar}}

\def\vinf{V_\infty}

\def\half{{\frac{1}{2}}}
\newcommand{\beq}{\begin{equation}}
\newcommand{\eeq}{\end{equation}}
\newcommand{\beqa}{\begin{eqnarray}}
\newcommand{\eeqa}{\end{eqnarray}}

\def\blankline{\par\vskip \baselineskip}
\def\hzd{\delta}

\date{\currenttime \today}

\begin{document}

\title[Analytic Dynamical Magnetosphere]
{
An `Analytic Dynamical Magnetosphere' formalism for X-ray and optical emission from slowly rotating magnetic massive stars
}

 \author[Owocki et al.]
{
\vbox{
Stanley P.\ Owocki$^1$\thanks{Email: owocki@udel.edu},
Asif ud-Doula$^2$,
Jon O. Sundqvist$^{3,4}$,  
Veronique Petit$^5$,  
David H.\ Cohen$^6$, and
Richard H.\ D. Townsend$^7$
}
\\ $^1$ Department of Physics and Astronomy, Bartol Research Institute,
 University of Delaware, Newark, DE 19716, USA
\\ $^2$ Penn State Worthington Scranton, Dunmore, PA 18512, USA
\\ $^3$  Centro de Astrobiolog{\'i}a, CSIC-INTA,\ Ctra. Torrej{\'o}n a Ajalvir km.4, 28850 Madrid, Spain
\\ $^4$ Instituut voor Sterrenkunde, KU Leuven, Celestijnenlaan 200D, 3001 Leuven, Belgium
\\ $^5$ Dept. of Physics and Space Sciences, Florida Institute of Technology, Melbourne, FL 32901
\\ $^6$ Department of Physics and Astronomy, Swarthmore College, 500 College Ave., Swarthmore, PA 19081, USA 
\\$^7$ Department of Astronomy, University of Wisconsin-Madison, 5534 Sterling Hall, 475 N Charter Street, Madison, WI 53706, USA
}

\maketitle

\begin{abstract}
Slowly rotating magnetic massive stars develop ``dynamical magnetospheres'' (DM's), characterized by trapping of stellar wind outflow in closed magnetic loops, shock heating from collision of the upflow from opposite loop footpoints,
and 
subsequent gravitational infall of radiatively cooled material.
In 2D and 3D magnetohydrodynamic (MHD) simulations the interplay among these three components is spatially complex and temporally variable, making it difficult to derive observational signatures and discern their overall scaling trends.
Within a simplified, steady-state analysis based on overall conservation principles, we present here an
``{\em analytic dynamical magnetosphere}'' (ADM) model that provides explicit formulae for density, temperature and flow speed in each of these three components -- wind outflow, hot post-shock gas, and cooled inflow -- as a function of colatitude and radius within the closed (presumed dipole) field lines of the magnetosphere.
We compare these scalings with time-averaged results from MHD simulations, and provide initial examples of application of this ADM model for deriving two key observational diagnostics, namely hydrogen H-$\alpha$ emission line profiles from the cooled infall, and X-ray emission from the hot post-shock gas.
We conclude with a discussion of key issues and advantages in applying this ADM formalism toward derivation of a broader set of observational diagnostics and scaling trends for massive stars with such dynamical magnetospheres.
\end{abstract}


\section{Introduction}

Hot luminous, massive stars of spectral type O and B have dense, high-speed, radiatively driven stellar winds
\citep{Castor75}.
In the subset ($\sim$10\%) of massive stars with strong ($> 1$kG), globally ordered (often significantly dipolar) magnetic fields
\citep{Petit13}, the trapping of the wind outflow by closed magnetic loops leads to the formation of a circumstellar {\em magnetosphere}.
For stars with moderate to rapid rotation -- such that the Keplerian corotation radius $R_K$ lies within the Alfv\'{e}n radius $R_A$ that characterizes the maximum height of closed loops --, the rotational support leads to formation of a ``{\em centrifugal magnetosphere}'' (CM), wherein the trapped wind material accumulates into a relatively dense, stable and long-lived rigidly rotating disk \citep{Townsend05}.

In contrast, for magnetic massive stars with slow rotation, and thus $R_K > R_A$, this trapped material  falls back to the star on a dynamical (freefall) timescale, representing then a ``{\em dynamical magnetosphere}'' (DM)  \citep{Sundqvist12c, Petit13}.
Because of the rotational spindown associated with angular momentum loss through the relatively strong, magnetized wind upflow from open field regions \citep{Uddoula09}, magnetic O-type stars are typically\footnote{The one exception is Plasket's star, for which the magnetic star likely has been spun up by mass accumulation from its binary companion 
\citep{Grunhut13}.} slow rotators, and so harbor DM's.
Among the magnetic B-stars, a significant fraction (about half) are also rotating slowly enough to have DM's 
\citep{Petit13}.

In such DM's the trapped wind upflow from opposite footpoints of closed loops collides near the loop apex, forming shocks that heat the gas to X-ray emitting temperatures;
as this gas radiatively cools, it falls back to the star as a gravitational downflow. 
2D and 3D magnetohydrodynamic (MHD) simulations of such DM's \citep{Uddoula02,Uddoula13} show a complex and variable interplay among all three components, and this makes it difficult to derive observational signatures and discern their overall scaling trends.

Applications of these MHD models have thus been limited to a few selected O-stars, using simplified radiative transfer methods to derive synthetic spectra for H$\alpha$ emission lines \citep{Sundqvist12c, Grunhut12, Uddoula13, Wade15} and ultra-violet wind resonance lines \citep{Marcolino13, Naze15}. 
These initial studies have provided strong support of the general DM concept;
however, the complexity of the time-dependent 3D structure, together with computational expense of the simulations, 
prohibits more systematic and detailed computations of synthetic observables across the 
electromagnetic spectrum, as well as application to larger samples of magnetic massive stars.

Similar arguments apply to X-ray spectral synthesis.
Detailed MHD simulations have been used to analyze the high spectral resolution X-ray observations available for a few 
selected OB stars
\citep{Petit15,Naze15}.
But for the analysis of the much larger number of stars with low-resolution X-ray data, an analytic model can capture key observable properties and trends.
Recently, \citet[][hereafter paper I]{Uddoula14} carried out an extensive MHD simulation study of radiative cooling of the hot post-shock gas in DM's, with a focus on deriving the resulting X-ray emission as a function of the density-dependent cooling efficiency.
When interpreted in terms of a simplified ``X-ray analytic dynamical magnetosphere" (XADM) analysis, this led to predicted scaling laws for variation of X-ray luminosity with the wind mass loss rate.
Subsequent application by \citet{Naze14} toward interpreting observationally inferred X-ray luminosities in a large sample of magnetic massive stars showed that, with some fixed factor adjustment to account for the limited ``duty cycle'' for X-ray production during the complex variations seen in MHD simulations, this XADM scaling matched quite well the observed trends for the subset of magnetic massive stars with slow enough rotation to have DM's.

The analysis here builds on this success to develop a more general ``analytic dynamical magnetosphere'' (ADM) formalism that now provides explicit formulae for the spatial variation of density and flow speed (as well as temperature for the hot gas) in all three components of the closed loop magnetosphere:
wind upflow, hot post-shock gas, and cooled downflow (section \ref{sec:adm}).
The overall ADM analysis here is guided and tested by comparison with time-averaged results from full MHD simulations of DM's (section \ref{sec:MHD}),
using the Alfv\'en radius $R_A$ in the MHD simulation to define a maximum closure radius $R_c$ of a dipole loop in the ADM model.

The inclusion of the cooled downflow now allows for modeling of optical emission lines like H-$\alpha$ (section \ref{sec:Balmer}).
Moreover, the description of the spatial distribution of both the hot and cool components allows an extension of the XADM analysis of paper I (section \ref{sec:xrays}),  for example by accounting for possible bound-free absorption of emitted X-rays by the cool wind and downflow \citep[see section \ref{sec:xray-abs} and][]{Petit15}.
We conclude (section \ref{sec:conclusions}) with a brief summary of key issues and advantages in applying this ADM formalism toward deriving a broader range of spectral diagnostics.
Appendices A and B  show how the ADM model can be used to derive both the  X-ray differential emission measure (DEM), as well as a shock-temperature distribution $p(T_s)$.
Appendix C gives background on the ADM scalings for H$\alpha$ emission.

\begin{figure}
\begin{center}
\vfill
\includegraphics[scale=0.48]{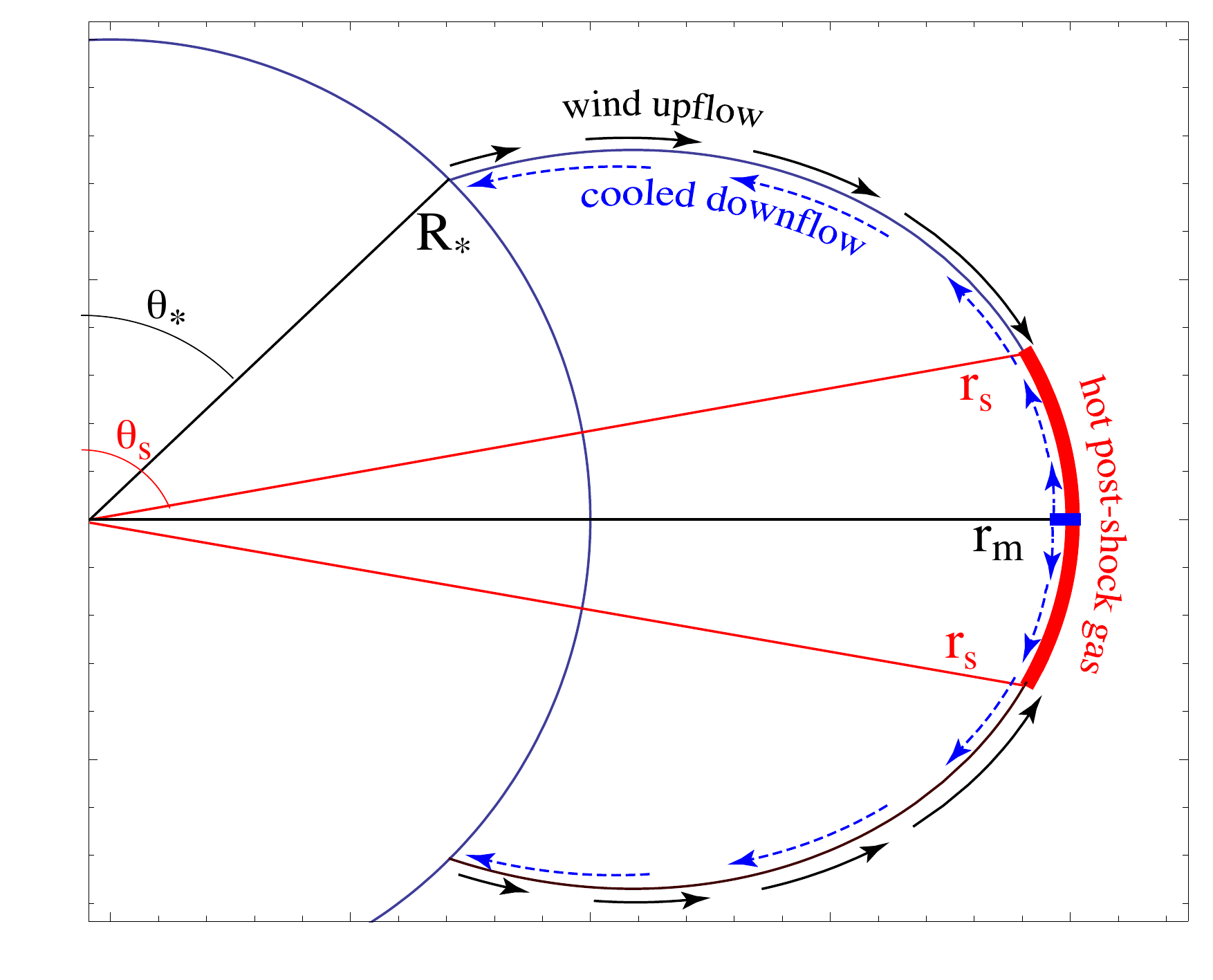}
\caption{
 Illustration of three components of material flow along a closed dipole loop line that intersects the stellar radius $\Rstar$ at a colatitude $\theta_\ast = \arccos \mustar$, and extends up to a maximum radius $r_m = \Rstar/ (1-\mustar^2)$ at the equatorial loop apex. 
 Wind upflow (black arrows) from a stellar surface footpoint meets wind material from the opposite footpoint at the loop apex.
 This results in a pair of reverse shocks with hot post-shock gas (red) extending back from the loop apex to shock fronts at co-latitudes $ \pi - \theta_s$ and $\theta_s (\equiv \arccos \mu_s )$, and radius $r_s = r_m (1-\mu_s^2)$.
 Cooled material at the loop apex (blue wedge) is pulled back by the stellar gravity, forming channels of cooled downflow (blue dashed arrows) back toward the star. 
}
\label{fig:dipgeom}
\end{center}
\end{figure}

\section{Analytic Dynamical Magnetosphere}
\label{sec:adm}

\subsection{Basic magnetic scalings for a star-centered dipole}
\label{sec:scdip}

The ADM formalism is based on an idealized, {\em steady-state} analysis of the mass flow along closed magnetic loops that are assumed to follow the individual lines of a simple 
dipole\footnote{For simplicity, this ignores the modest outward stretching  of closed loop lines by the dynamical pressure of the trapped wind upflow, e.g. as seen in the MHD simulations shown in figure \ref{fig:adm-vs-mhd}.} 
magnetic field, taken here to be centered at the  central radius $r=0$ of the underlying star.

Following figure \ref{fig:dipgeom},
if we specify the colatitude $\theta$ from the north dipole axis by $\mu = \cos \theta$, then a given dipole loop line that intersects the stellar radius $\Rstar$ at $\mu=\mustar$ extends over a band $-\mustar <  \mu < \mustar$ about the equator, with radius variation given by 
\beq
\frac{r(\mu,\mustar)}{\Rstar } = \frac{1-\mu^2}{1-\mustar^2} ~~; ~~ |\mu| < \mustar
\, ,
\label{eq:rmu}
\eeq
and with the maximum radius at the loop apex on the magnetic equator,
\beq
r_m (\mustar) \equiv  r(0,\mustar ) =  \frac{\Rstar}{1-\mustar^2} 
\, .
\label{eq:rm}
\eeq
For any position $\{r,\mu\}$ within the magnetosphere, the magnitude of the field follows the dipole scaling,
\beq
\frac{B(r,\mu)}
{B (\Rstar,\mustar)}
=
 \left ( \frac{\Rstar}{r} \right )^3 \,
\left ( \frac{ 1 + 3 \mu^2 }{1 + 3 \mustar^2 } \right )^{1/2}
\, .
\label{eq:bmagrmu}
\eeq
For such a dipole, the direction of the field only depends on $\mu$, given by the unit vector
\beq
\hat{\bf b} (\mu) \equiv \frac{\bf B}{B} = \frac{2 \mu \hat{\bf r} + \sqrt{1-\mu^2} ~ \hat{\bf \theta }}{\sqrt{1+3\mu^2}}
\, ,
\label{eq:bhatrmu}
\eeq
where $\hat{\bf r}$ and $\hat{\bf \theta}$ are unit vectors in the radial and latitudinal directions.

The nearly full ionization of circumstellar material around hot, OB stars implies a high conductivity and thus broad applicability of the {\em frozen flux} theorem of ideal MHD. 
In the context of the present model of a rigid, closed dipole field line, this means any material remains  always locked onto its given field line, with flow velocity ${\bf v}$ parallel to the local field direction $\hat{\bf b}$.
In terms of the local density $\rho$, the mass flux $\rho {\bf v}$  thus has a divergence given by,
\beq
\nabla \cdot (\rho {\bf v} ) = \nabla \cdot (\rho v {\bf B}/B)
 = {\bf B} \cdot \nabla (\rho v/B ) = B \frac{\partial (\rho v/B)}{\partial b}
\eeq
where the third equality uses $\nabla \cdot {\bf B} = 0$, and  the last equality defines a coordinate distance $b$ along the field line, with $\hat{\bf b} \cdot \nabla \equiv \partial/\partial b$.
This means that in any steady-state flow, wherein mass conservation requires $\nabla \cdot (\rho {\bf v}) = 0$,  the local mass flux density scales with the field strength, so that $\rho v / B =$~{\em constant} along the flow.
For a given steady input mass flux from the surface, the spatial variation of density $\rho$ can thus be derived from knowledge of the flow speed $v$ (or vice versa), in terms of the known spatial variation of field strength  $B$ from (\ref{eq:bmagrmu}).

\subsection{3-Component model for mass trapped in a closed dipole loop}
\label{sec:3comp}

In full MHD simulations (see paper I), the actual flow along such closed loops is  spatially structured and time-dependent, with any given loop alternating between variable intervals and regions of upflow and downflow. As demonstrated below, however, the overall conservation principles mean that, when averaged over time, and/or over many separate loops with complex structure, key characteristics from 2D and 3D MHD simulations can be relatively well characterized by an idealized ADM model that assumes
{\em each loop line can simultaneously support two oppositely directed, steady-state flows}. 

Specifically, as illustrated in figure \ref{fig:dipgeom}, this ADM analysis distinguishes three distinct components of material flow within the loop:
\begin{enumerate}
\item{\em Wind upflow}

\item{\em Hot post-shock gas}

\item{\em Cooled downflow}

\end{enumerate}

Mass is fed into the loop by the {\em wind upflow}, driven by the radiative flux from the underlying star.
Relative to the surface mass flux $\Mdot_{B=0}/4\pi \Rstar^2$ for the non-magnetic case, 
the mass flux $\mdot_b$ fed into a loop with footpoint $\{\Rstar, \mustar\}$ scales as 
\citep{Owocki04c} 
\beq
 \mdot_b (\mustar)
= \mu_B (\mustar)  = \frac{2 \mustar}{\sqrt{1+3 \mustar^2}}
\, ,
\label{eq:mds}
\eeq
where $\mu_B$ is the radial projection cosine of the local surface field\footnote{The {\em radial} mass flux scales with $\mu_B^2$, but the flux along the field line scales linearly with $\mu_B$.}, 
and the second equality applies equation \ (\ref{eq:bhatrmu}) for a standard dipole.

The flow speed $v$ along the loop should in principle be computed from solution of the acceleration from radiative driving, accounting for the curvature, tilt and areal divergence along the loop; 
but to maintain analytic tractability, our ADM analysis simply assumes this speed follows a canonical ``beta'' velocity law,
\beq
\frac{v(r)}{\vinf } \equiv w(r) = (1-\Rstar/r)^\beta 
\, ,
\label{eq:betalaw}
\eeq
where the maximum speed $\vinf = v(r \rightarrow \infty )$ is given here by the expected terminal speed of a corresponding unmagnetized wind, and we assume the simple case with $\beta=1$.

As the upflow from a loop footpoint approaches the loop apex at radius $r_m$, where the scaled speed reaches its maximum loop-specific value $w_m = 1-\Rstar/r_m = \mustar^2$, the collision with flow from the opposite footpoint induces a pair of reverse shocks, 
one on each side of the loop apex. 
This converts the wind kinetic energy into heat, resulting then in the {\em hot post-shock gas} that extends away from the loop apex by a distance set by the post-shock cooling length $\ell_c$.
For lower luminosity stars with smaller mass loss rate $\Mdot_{B=0}$ and thus a lower-density wind upflow, the cooling length can become comparable to the loop apex radius, $\ell_c \lesssim r_m$. 
As discussed in paper I,
 this forces a ``{\em shock retreat}'' to a lower shock radius $r_s < r_m$, with thus a slower scaled  shock speed $w_s = w(r_s) = 1 - \Rstar/r_s$ and so a cooler post-shock temperature $T_s \sim w_s^2$.
 Using the dipole shock-retreat analysis in Appendix B of paper I (recapitulated in section \ref{sec:hpsg} below),
 figure \ref{fig:shocklines} illustrates the progressive retreat of the shock with increased cooling length, as characterized by a cooling parameter $\chi_\infty$, given below in equation (\ref{eq:chiinf}).
 
\begin{figure}
\begin{center}
\vfill
\includegraphics[scale=0.42]{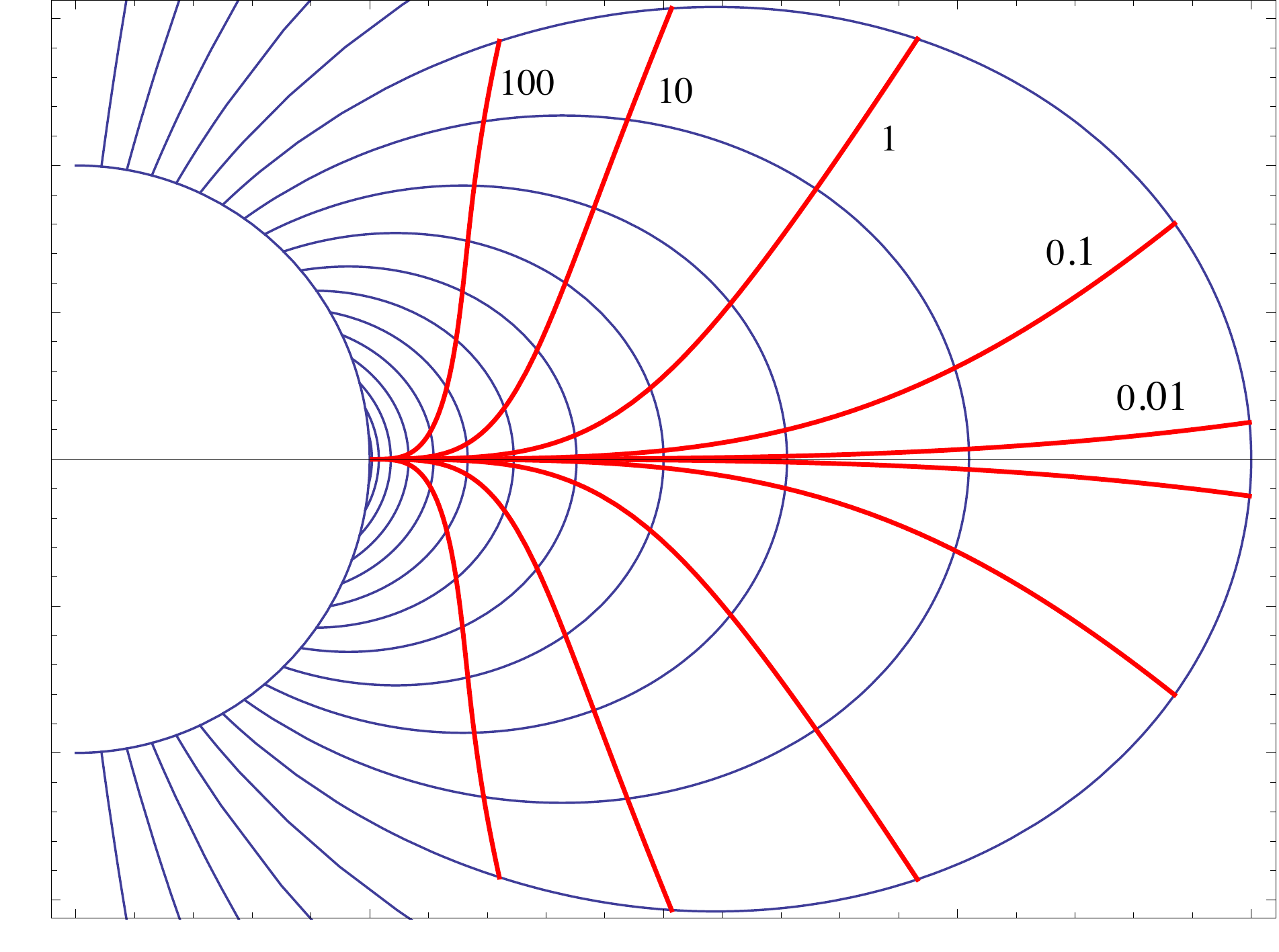}
\caption
{Dipole magnetic field lines (blue curves) along with the retreated shock locations (red curves), labeled with cooling parameters ranging from $\chi_\infty = 0.01$ (closest to magnetic equator) to $\chi_\infty = 100$ (furthest from equator) in steps of 1 dex
}
\label{fig:shocklines}
\end{center}
\end{figure}

Starting from this immediate post-shock temperature $T_s$ of order many MK, radiative cooling in the post-shock region $r> r_s$ causes a decrease in temperature along the loop, eventually reaching near the loop apex the same stellar photoionization equilibrium temperature of the wind upflow, typically on the order of the stellar effective temperature of a few $10$\,kK.
The subsonic nature of the post-shock flow means this cooling layer is almost isobaric, with the nearly constant pressure $P \sim \rho T$  implying a strong increase in density $\rho$ as the temperature $T$ declines.

Much like a ball balanced at the top of a hill,  the convex upward nature of the magnetic loop near its apex means such dense, 
cooled material is gravitationally unstable, and so begins a gravitational free-fall along the loop to form the {\em cooled downflow} component of the DM.
The strong compression, and relatively slow infall speed, means the density is much higher than in the wind upflow, 
and the several $10$kK equilibrium temperature means it can efficiently radiate in hydrogen emission line series, leading then to prominent optical emission in lines like H$\alpha$ \citep{Howarth07, Sundqvist12c} that are much stronger than from the wind.
Moreover it is cool enough that there is a significant X-ray bound-free opacity from abundant, partially ionized heavy elements like CNO and Fe, which can thus play a role in attenuating the X-rays emitted from the hot post-shock gas
\citep{Petit15}.

As noted, in 2D and 3D MHD simulation of DM's \citep[][paper I]{Uddoula13}, these three components of wind upflow, hot post-shock gas, and cooled downflow become mixed together in complex, highly variable combinations, with 3D models showing structure down to scales of $\Rstar/100$ or less, such that even loop lines separated by such a small scale can be in markedly different phases for the cycle of shock-build-up, cooling, and then infall \citep{Uddoula13}.

But as we demonstrate in section \ref{sec:MHD} below, if one takes a suitably long time average, covering many such infall cycles, then these stochastic variations become smoothed out, leaving a distinctive and organized large-scale spatial distribution. 
This can be well characterized by a superposition of the density, velocity and temperature of the wind upflow, hot post-shock gas, and cooled downflow.
The next subsections quantify this in terms of relatively simple ADM scalings based on quasi-steady-state conservation applied to the material from each component.

\subsubsection{Wind upflow}
\label{sec:windupflow}

The conditions in the wind upflow  are quite straightforward to specify. 
As noted, the speed is assumed to follow the $\beta=1$  law given in equation (\ref{eq:betalaw}), 
\beq
v_w (r)  = \vinf w(r) = \vinf (1-\Rstar/r)
\, .
\label{eq:vw}
\eeq
The density then follows from steady-state mass continuity, with base mass flux (\ref{eq:mds}), and accounting for the dipole area divergence,
\beq
\frac{\rho_w (r,\mu) }{\rho_{w\ast}} =
\frac{\mdot_b (\mustar) }{w(r)} \,
\frac{B(r,\mu)}{\Bstar (\mustar )} \, ,
\label{eq:rhow1}
\eeq
where 
$\rho_{w\ast} \equiv \Mdot_{\rm B=0}/4 \pi \Rstar^2 \vinf$ is a characteristic density for an unmagnetized wind with mass loss rate
$\Mdot_{\rm B=0}$ and terminal speed $\vinf$. 

Using equations (\ref{eq:rmu}),  (\ref{eq:bmagrmu}), (\ref{eq:mds}), and (\ref{eq:vw}),
we can rewrite this in a form that 
explicitly shows the dependence on radius and co-latitude,
\beq
\frac{\rho_w (r,\mu) }{\rho_{w\ast}} 
=
\frac{\sqrt{r/\Rstar - 1 + \mu^2} \, \sqrt{1 + 3 \mu^2} }
{(r/\Rstar - 1) \, (4r/\Rstar -3  + 3 \mu^2)} \,
\left ( \frac{\Rstar}{r} \right )^{3/2} .
\label{eq:rhow2}
\eeq

Finally, the temperature of the wind upflow is expected to be of order the stellar effective temperature, but its actual value is not relevant to the ADM model, so long as the upflow is sufficiently supersonic to justify use of the strong-shock scaling in modeling the post-shock flow, as discussed next.
\begin{figure*}
\begin{center}
\vfill
\includegraphics[scale=0.585]{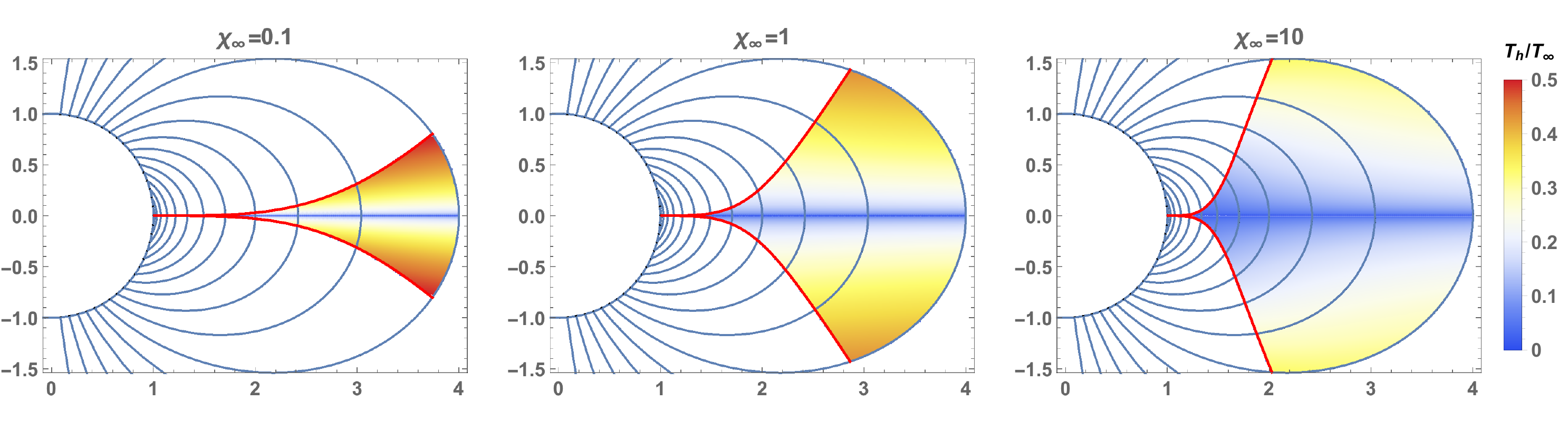}
\vskip 0.2cm
\includegraphics[scale=0.575]{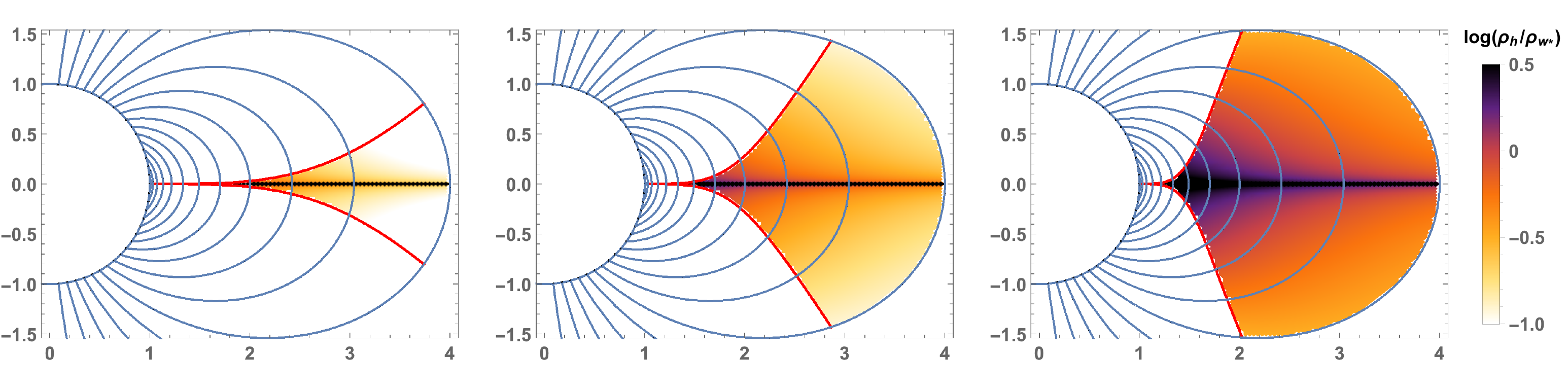}
\vskip 0.2cm
\caption{Mosaic of ADM hot-component properties. 
The left, center, and right columns are  for cooling parameters $\chi_\infty = 0.1$, 1, and 10, respectively.
The upper row shows hot post-shock temperature, scaled by maximum wind-shock temperature,  $T_h/T_\infty$.
The bottom row shows log of the density in the post-shock hot gas ($\rho_h$), scaled by the characteristic 
wind density $\rho_{w\ast} \equiv \Mdot_{B=0}/4 \pi \vinf \Rstar^2$.
}
\label{fig:hall}
\end{center}
\end{figure*}

\begin{figure*}
\begin{center}
\includegraphics[scale=0.57]{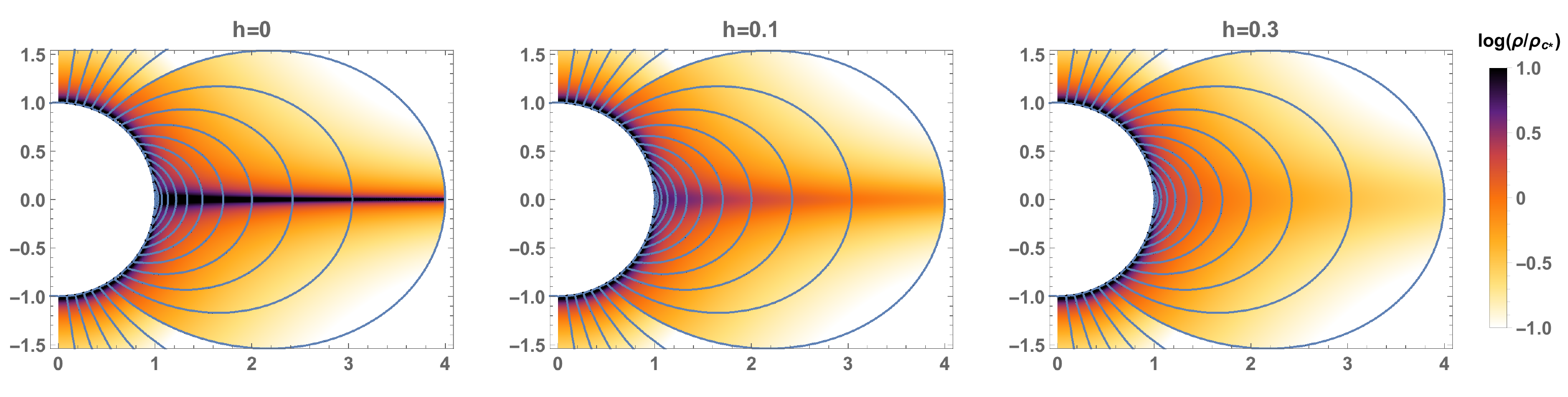}
\caption{Log of the density of cool material from both the cooled downflow, $\rho_c$ and the wind upflow $\rho_w$, 
scaled in units of $\rho_{c\ast}$, and assuming $v_\infty/v_{\rm e}=3$.
The 3 panels show results for apex smoothing lengths $\hzd/\Rstar=$ 0, 0.1, and 0.3 (left, center, right).
}
\label{fig:dch}
\end{center}
\end{figure*}

\subsubsection{Hot post-shock gas}
\label{sec:hpsg}

Let us next derive conditions in the hot post-shock gas in the region between the shock radius and loop apex, $r_s < r < r_m$.
Our analysis builds on and extends the X-ray analytic dynamical magnetosphere (XADM) formalism developed in paper I, 
deriving now explicit expressions for the spatial variation of the temperature in this post-shock cooling layer.

We begin with the equation for advective change of the specific enthalpy $(5/2) kT$ due to radiative cooling, characterized by a radiative cooling function $\Lambda $, which we write in a mass-weighted form $\Lambda_m = \Lambda/\mu_p \mu_e$, with $\mu_p$ and $\mu_e$ respectively the mean mass per proton and per electron
(see section 2 of \citet{Kee14}, and section 2.5 and Appendix B of paper I),
\beq
\frac{5k}{2} \, v \frac{d T}{db} = - \mubar \rho \Lambda_m
\, ,
\label{eq:dTdb1}
\eeq
where $\mubar$ is the mean molecular weight, and 
the temperature derivative is with respect to the field line coordinate $b$.
The nearly constant pressure $P \sim \rho T$ of this post-shock layer implies a strong increase in density $\rho$ as temperaure $T$ declines.

Using steady mass flux conservation $\rho v \sim \def\Rstar{R_{\ast}}B$ and this near constancy of the pressure $P \sim \rho T$, we can eliminate both the density $\rho$ and the flow speed $v$ in favor of the temperature $T$. 
Following the analysis in Appendix B of paper I,
we can then integrate (\ref{eq:dTdb1}) to obtain an implicit solution for the temperature decline from the post-shock value $T_s$ to the much lower (near zero) temperature at the loop apex,
\beq
1- \left ( \frac{T}{T_s} \right )^3 = 
\frac{3}{\chi_\infty} \, \frac{\mdot_b}{w_s^4} \, \frac{B_s^2}{\Bstar B_m} \, \frac{r_m}{\Rstar} \, 
\int_{b(r_{s})}^{b(r)} \frac{B_m}{B} \, \frac{db}{r_m}
\, ;
\label{eq:Timplicit}
\eeq
here $B_m \equiv B(r_m)$, and
$\mdot_b$ accounts for the mass loss weighting for a given field-line flow tube, defined as a fraction of the spherical mass loss $\Mdot_{B=0}$ used in the definition of the cooling parameter (taken from equation (25) of paper I),
\beq
\chi_\infty  \equiv  \frac{15 \pi  }{128} \, \frac{V_\infty^4 \Rstar} { \Mdot_{B=0}  \Lambda_m  }
\approx
0.034  \, \frac{V_{8}^4 \, R_{12} }{\Mdot_{-6}}
\, ,
\label{eq:chiinf}
\eeq
with scaled values 
$R_{12} \equiv \Rstar/10^{12} \,$cm, 
$V_{8} \equiv \vinf/10^8$\,cm/s,
and 
$\Mdot_{-6} \equiv \Mdot_{B=0}/10^{-6} \Msun$/yr.

Following eqns.\ (B11)-(B14) of paper I,
the path integral of the field strength along the loop can be evaluated as
\beq
\int_{b(r_{s})}^{b(r)} \frac{B_m}{B} \, \frac{db}{r_m} 
= g(\mu_s) - g(\mu(r)) 
 \, ,
 \label{eq:intg}
\eeq
where
\beq
g(\mu) \equiv 
\int_{0}^{|\mu|}  \left ( 1- \mu^2 \right )^3 \, d\mu =
\left | \mu - \mu^3 + \frac{3 \mu^5}{5} - \frac{\mu^7}{7} \right |
\, ,
\label{eq:gmu}
\eeq
with the absolute value operations ensuring that $g( \mu)$ remains positive even in the lower hemisphere, where $\mu < 0$.
Applying the boundary condition that the temperature nearly vanishes at the loop apex, and so formally taking $T(r_m) \approx 0$, we find, since $g(0)=0$, that the spatial variation of temperature in this hot component of the ADM can be written as
\beq
{\tilde T}_h (r,\mu) = 
T_s
\left  [ \frac{g(\mu )} {g(\mu_s)} \right ]^{1/3} ~~ ; ~~ r_s < r < r_m
\, ,
\label{eq:Thtilde}
\eeq
where, for the assumed case of a highly supersonic outflow yielding a strong shock,  the immediate post-shock 
temperature $T_s = T_\infty w_s^2$, with
\beq
T_\infty = \frac{3}{16} \frac{\mubar \vinf^2}{k} \approx 14 \, {\rm MK} ~ V_8^2 
\approx 1.2 \, {\rm keV} ~ V_8^2 
\, .
\label{eq:Tinfdef}
\eeq
In practice, to account for the effects of photoionization heating by the underlying star, we do not allow the temperature to fall below the stellar effective temperature, 
\beq
T_h(r,\mu) = \max[{\tilde T}_h (r,\mu), T_{\rm eff} ]
\, ,
\label{eq:Thr}
\eeq
where here we assume a typical hot-star value $T_{\rm eff} = $30,000\,K.

The shock radius $r_s = r_m (1-\mu_s^2)$ is obtained by solving a transcendental equation for the dipole shock retreat, as derived in equation (B16) of paper I, 
\beq
g(\mu_s)
=   \frac{\chi_\infty}{6\mustar} 
 \, \frac{1+3\mustar^2}{1+3\mu_s^2} \,  \left( \frac{w_s r_s}{r_m} \right )^4  \left( \frac{r_s}{\Rstar} \right )^2 
 \, .
\label{eq:wsdip}
\eeq
Figure 3 of paper I plots the variation of the scaled shock speed $w_s$ vs. cooling parameter $\chi_\infty$, for various loop apex speeds $w_m$.
The red lines in figure \ref{fig:shocklines} here illustrate the progressive spatial retreat of the shock away from the loop apex at the magnetic equator as the radiative cooling parameter is increased from a small ($\chi_\infty = 0.01$) to large ($\chi_\infty = 100$) values in steps of 1 dex.

Let us next obtain the spatial variation of the density for this hot, post-shock region.
For a strong shock, the density of the immediate post-shock gas is simply a factor four times the incoming wind density at the shock, $\rho_s =4 \rho_w (r_s)$.
Since the pressure $P \sim \rho T$ is nearly constant in this subsonic post-shock cooling layer,
we can write the spatial variation of density of the hot gas as
\beq
\rho_h (r,\mu) = 4 \rho_w (r_s,\mu_s) \,  
\frac{T_s}{T_h(r,\mu)} \,
~~ ; ~~ r_s < r < r_m \, ,
\label{eq:rhohr}
\eeq
where $T_h$ is obtained from equations (\ref{eq:Thr}) and (\ref{eq:Thtilde}).

Using mass continuity, we can also readily derive the spatial variation of the post-shock flow speed.
For a strong shock, the immediate post-shock speed is just a quarter of the incoming wind speed, $v_s = v_w (r_s)/4$, while the spatial variation is given by
\beq
v_h (r,\mu) = \frac{w_s \vinf}{4} \,  
\frac{T_h(r,\mu)}{T_s} \, 
\frac{B(r,\mu)}{B_s}
 ~~ ; ~~ r_s < r < r_m
\, ,
\label{eq:vhr}
\eeq
where the field magnitude $B(r,\mu)$ is given by (\ref{eq:bmagrmu}).

The top two rows of figure \ref{fig:hall} plot results for the spatial variation of $T_h$ (top) and $\log \rho$ (middle), for the 3 cooling parameter values $\chi_\infty =$0.1, 1, and 10 (left, middle, and right columns).

\subsubsection{Cooled downflow}
\label{sec:cooldownflow}

Once this hot post-shock gas cools, the stellar gravity pulls the cooled material back to the star, accelerating from near zero speed at the loop apex at $r_m$, into a {\em cooled downflow} along the loop.
For effective\footnote{``Effective'' here means reduced to account for the reduction in effective gravity from electron scattering. Because the high density means most lines will be saturated, we ignore the effect of line-opacity in reducing gravity.} 
stellar mass $\Mstar$ and radius $\Rstar$, and so surface escape speed $v_{\rm e} \equiv \sqrt{2 G\Mstar/\Rstar}$,
conservation of gravitational + kinetic energy gives for the cold gas downflow speed,
\beq
v_c(r,\mu) = 
v_{\rm e} \, \sqrt{  \frac{\Rstar}{r}  - \frac{\Rstar}{r_m} } = 
v_{\rm e} \, | \mu | \, \sqrt{\frac{\Rstar}{r}} 
\, .
\label{eq:vcrmu}
\eeq

Using mass conservation, the associated density $\rho_c$ can be computed in a completely analogous way as the wind upflow density $\rho_w$ in equation (\ref{eq:rhow1}), but now replacing the wind upflow speed $v_w$ with the cooled downflow speed $v_c$ , 
\beq
\frac{\rho_c (r,\mu)  }{\rho_{c \ast}}
 =
 \mdot_b  (\mustar ) \,
\frac{\sqrt{r/\Rstar}}{ \mu_\hzd} \, 
 \frac{B(r,\mu)}{B (\Rstar,\mustar )}
 \,  ,
\label{eq:rhoc1}
\eeq
where $\rho_{c \ast} \equiv \Mdot_{B=0}/4 \pi \Rstar^2 v_{\rm e} = \rho_{w \ast} \vinf/v_{\rm e}$
is a characteristic density for the downflow.
To account for the fact that this infall is typically initiated from some finite length $\hzd$ away from the exact loop apex,
in the denominator here we have replaced the $\mu$ factor in equation (\ref{eq:vcrmu}) for the speed $v_c$ with 
\beq
\mu_\hzd \equiv \sqrt{\mu^2 + \hzd^2/r^2}
\, ,
\label{eq:muh}
\eeq
which has the effect of smoothing the density near the magnetic equator over the scale $\hzd$.
Using this and equations (\ref{eq:rmu}),  (\ref{eq:bmagrmu}), and (\ref{eq:mds}),
we can rewrite (\ref{eq:rhoc1}) in a form that explicitly shows the dependence on radius and co-latitude,
\beq
\frac{\rho_c (r,\mu)  }{\rho_{c \ast}} =
 \frac{ \sqrt{ r/\Rstar - 1 + \mu^2} \, \sqrt{1 + 3 \mu^2} }
{\sqrt{\mu^2 + \hzd^2/r^2} \, (4r/\Rstar -3  + 3 \mu^2)} \,
\left ( \frac{\Rstar}{r} \right )^{2} 
 \, .
\label{eq:rhoc2}
\eeq
The value of this apex smoothing length $\hzd$ can be set based on results from MHD simulations,
or derived from comparison to observations (see section~\ref{sec:Balmer}).

\begin{figure*}
\begin{center}
\includegraphics[scale=0.45]{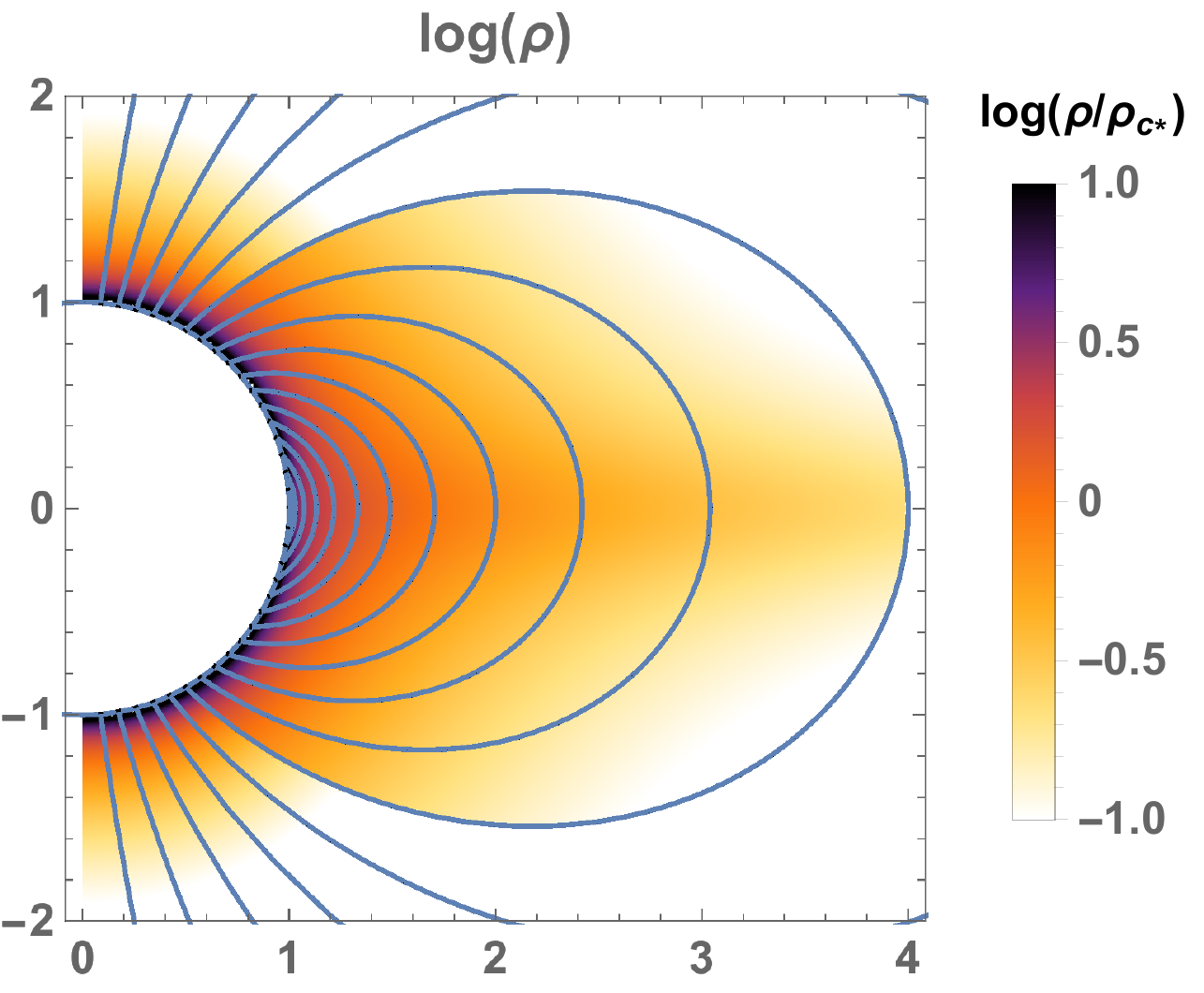}
\includegraphics[scale=0.46]{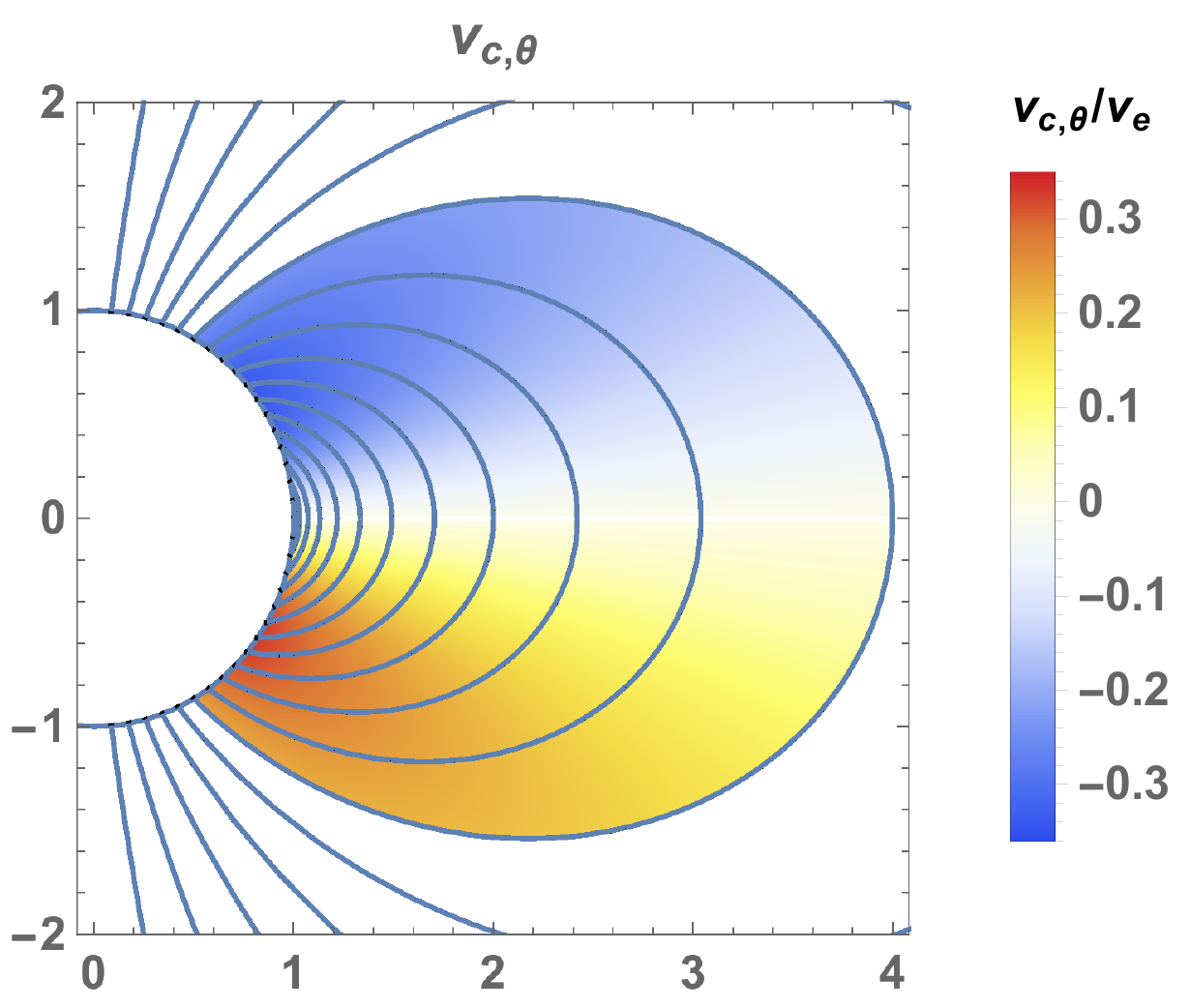}
\includegraphics[scale=0.44]{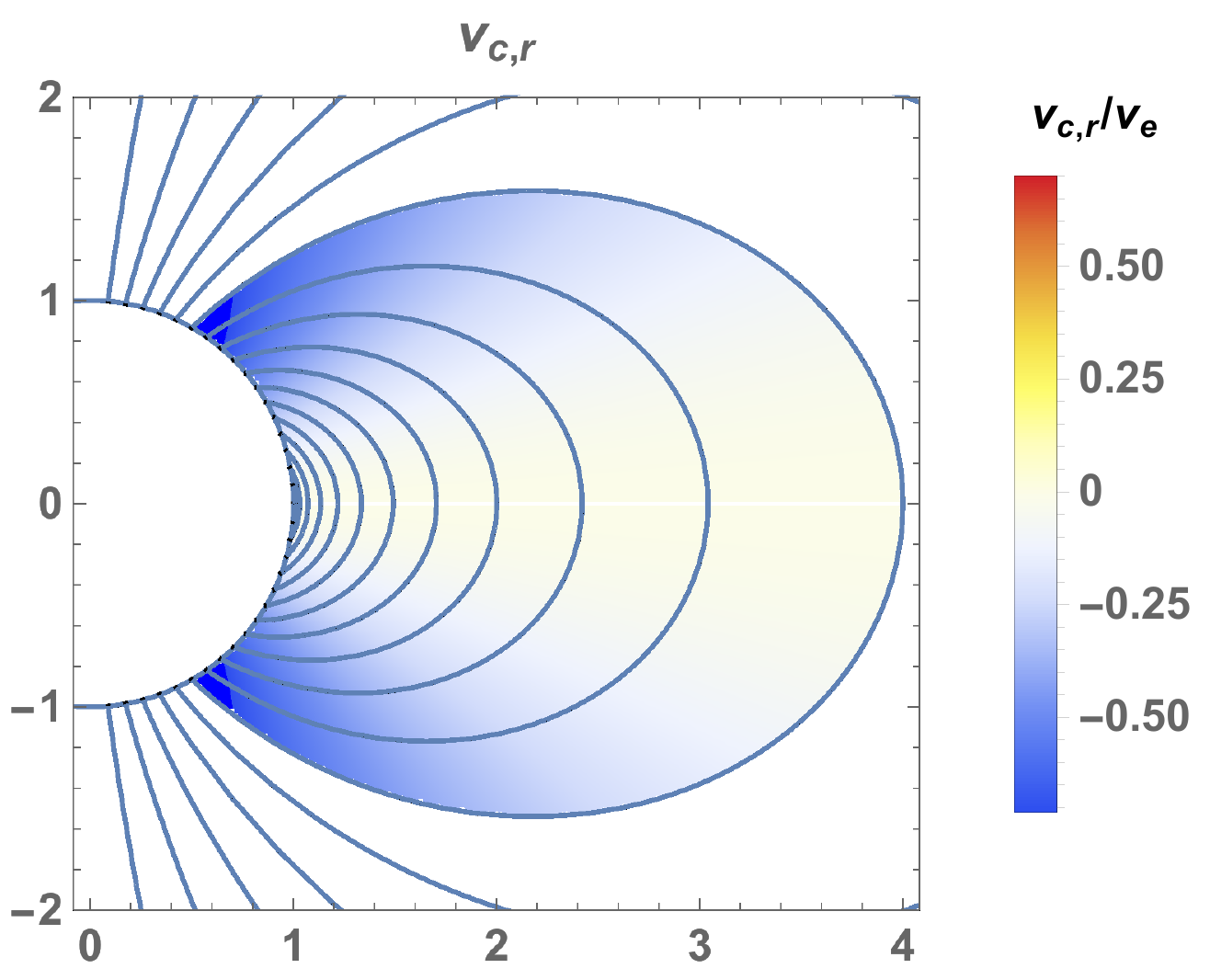}
\includegraphics[scale=0.46]{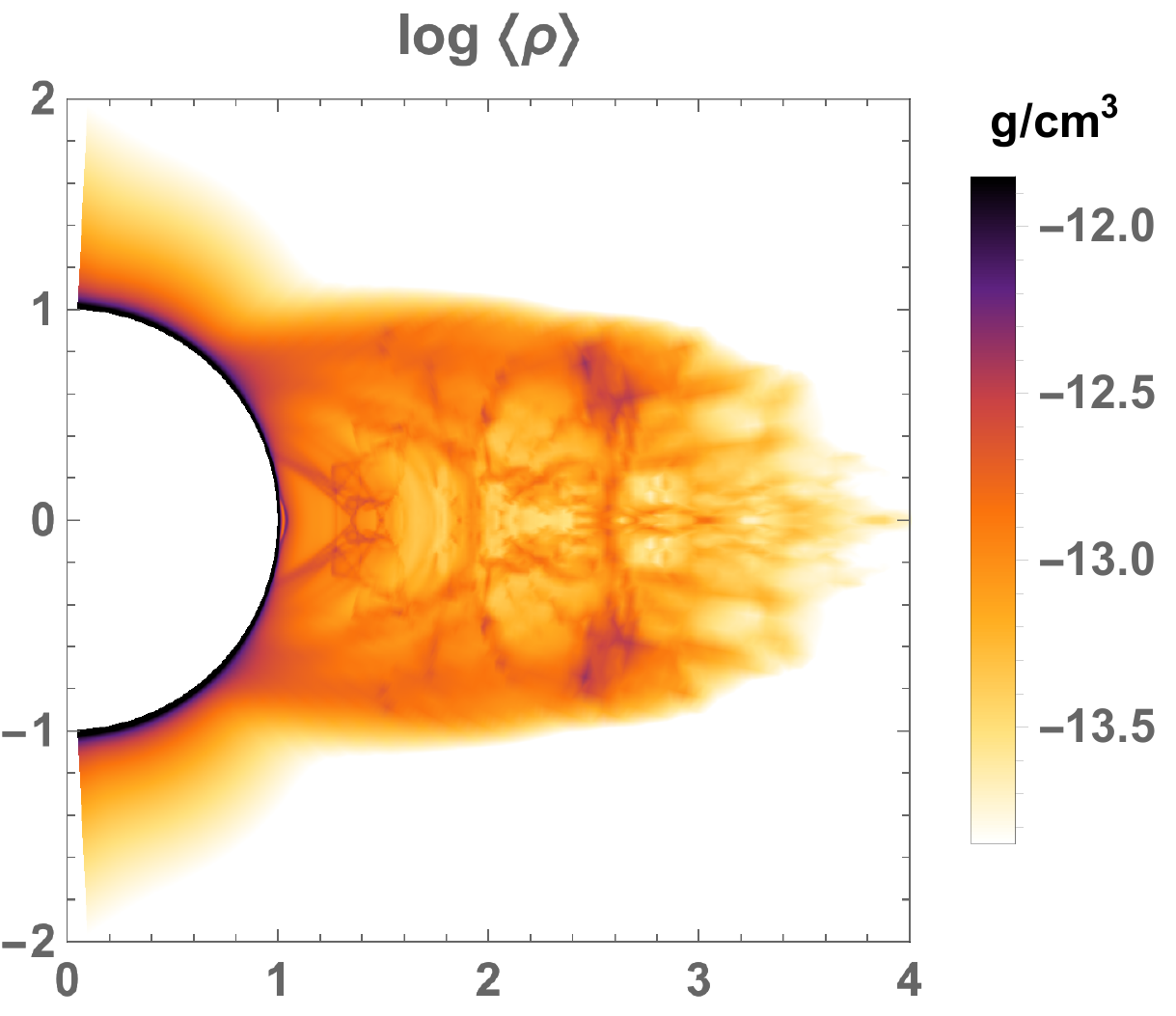}
~~
\includegraphics[scale=0.46]{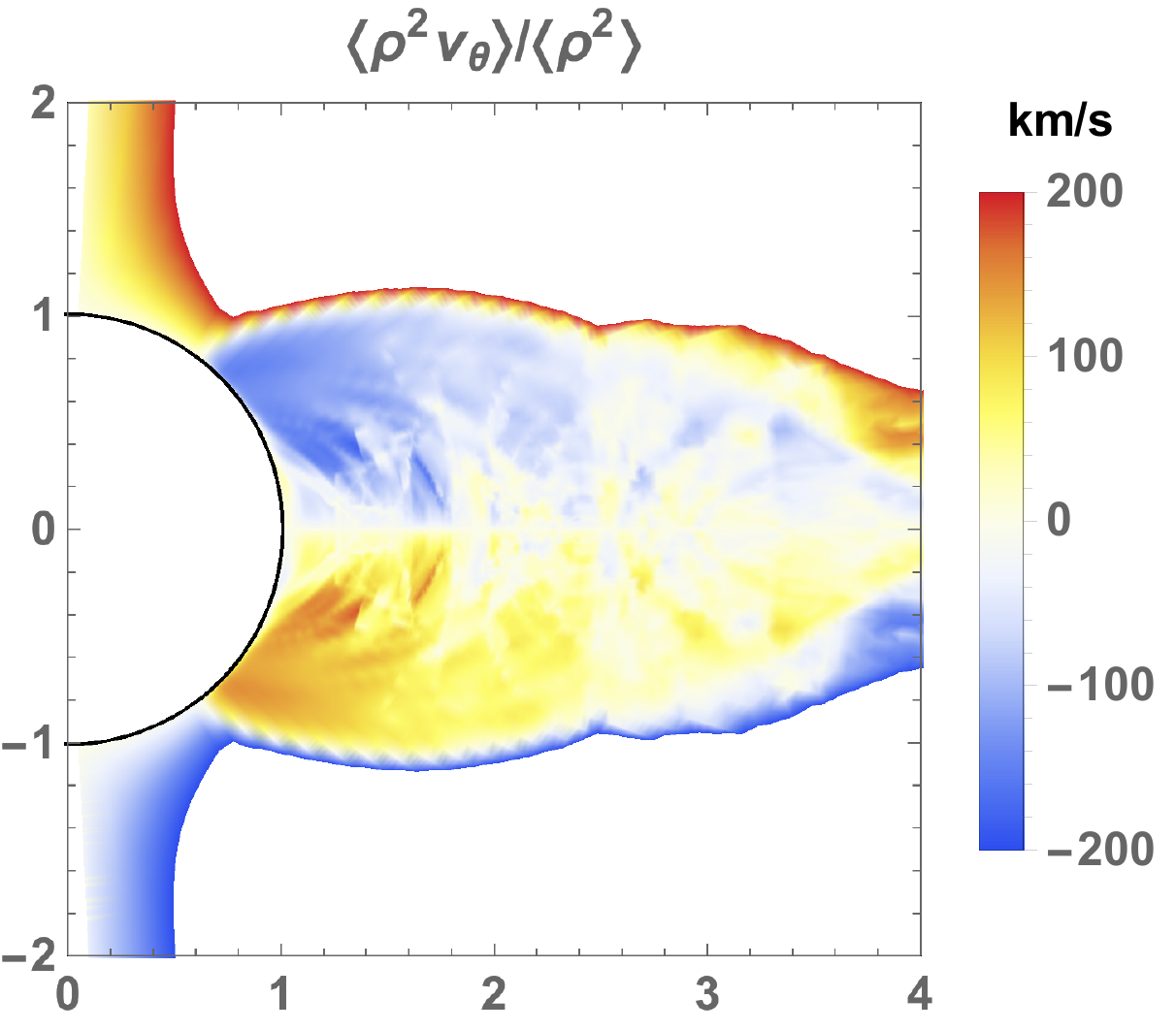}
~~s
\includegraphics[scale=0.46]{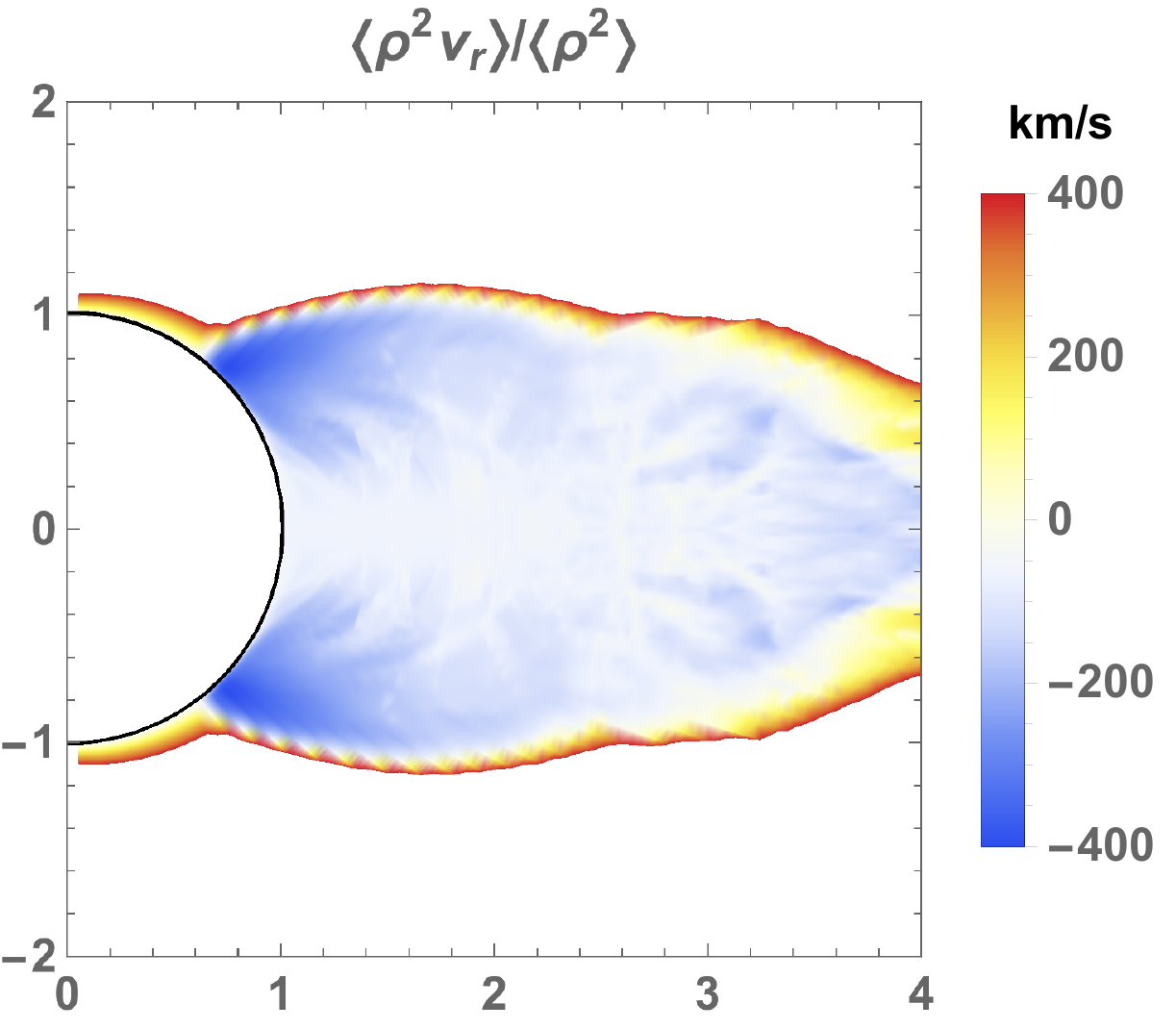}
\caption{
{\em Top row}: 
For an ADM model with maximum closure radius $R_c = 4 \Rstar$, the top left panel shows log density (scaled by $\rho_{c*}$), 
given by the wind outflow component for loops with apex radii $r_m > R_c$, and by the  cooled downflow component (with $\hzd/\Rstar=0.3$) for loops with $r_m \le R_c$.
The middle and right panels show the cooled downflow speed in the latitudinal ($v_{c,\theta}$) and radial ($v_{c,r}$ directions, scaled by the stellar surface escape speed $v_{\rm e}$.
{\em Bottom row}:
Corresponding time averages for MHD simulations, 
starting in the lower left with the mean density $ \left < \rho \right > $.
To reflect the scaling of line emission, we show
and the density-squared-weighted velocity ($ \left < \rho^2  {\bf v} \right >/\left < \rho^2 \right >$)
in the latitudinal (middle) and radial (right) directions.
The MHD simulations use parameters that give an Alfv\'{e}n radius that is approximately equal to the maximum closure radius of the ADM model, $R_A \approx 4 \Rstar = R_c$.
In the MHD results, the appearance of the wind speed in open regions is suppressed by truncating values outside the given colorbar ranges to white.
To suppress the impact of stochastic, asymmetric north-south variations in these 2D MHD simulations, the data has been north-south symmetrized to provide a clearer correspondence to the inherent symmetry of the ADM model.
In the middle and right lower panels for the MHD simulations, we have set to white any flow that is beyond the quoted colorbar ranges;
this has the effect of clearly delineating between open and close field regions, thus allowing a clearer comparison with corresponding ADM results that just show closed-loop infall.
The colorbar labels are in CGS units for the MHD simulation; quantitative comparison to  the ADM can be done by scaling these with the associated values $\rho_{c\ast} \approx 1.5 \times 10^{-13}$\,g\,cm$^{-3}$ for characteristic wind density, and $v_{\rm e} \approx $\,700\,km\,s$^{-1}$.
}
\label{fig:adm-vs-mhd}
\end{center}
\end{figure*}

The colorplot in figure \ref{fig:dch} shows the spatial variation of this cooled downflow density, $\log \rho_c/\rho_{c \ast}$, for 3 selected values of the smoothing length,
$\hzd/\Rstar =$ 0, 0.1, and 0.3 (left, center, right).
The top row of figure \ref{fig:adm-vs-mhd} shows the density for both the wind and cooled downflow for the case $\hzd/\Rstar=0.3$ (left panel), along with latitudinal and radial components of the downflow velocity (middle and right panels).
The lower panel shows corresponding time-averages from full MHD simulations, as discussed further in section \ref{sec:adm-mhd-cd}.

Finally, as with the wind upflow, the temperature of the cooled downflow is expected to be of order the stellar effective temperature, but its actual value and  spatial variation is not set in this ADM formalism, and so can be modeled separately in the context of its relevance to each specific diagnostic.

\subsection{Summary scalings for ADM components}
\label{sec:sclsum}

To facilitate application of this ADM model in deriving observational diagnosts, let us collect here the scaling relations for temperature, speed, and density in each of the three model components.
\blankline
\noindent
For the wind upflow (section \ref{sec:windupflow}):

$v_w$ is given by equation (\ref{eq:vw});

$\rho_w$ is given by equation (\ref{eq:rhow2}).

\blankline
\noindent
For the hot component (section \ref{sec:hpsg}):

$T_h$ is given by equations (\ref{eq:Thr}) and (\ref{eq:Thtilde});

$v_h$ is given by equation (\ref{eq:vhr});

$\rho_h$ is given by equation (\ref{eq:rhohr}),

\noindent
with all 3 using the auxiliary relations (\ref{eq:gmu}), for the field line geometry function $g(\mu$), and (\ref{eq:wsdip}), for shock location, $\{r_s,\mu_s\}$.

\blankline
\noindent
For the cooled downflow (section \ref{sec:cooldownflow}):

$v_c$ is given by equation (\ref{eq:vcrmu});

$\rho_c$ is given by equation (\ref{eq:rhoc2}).

\blankline
\noindent
In the plots here, 
lengths are in stellar radii $\Rstar$, 
velocities are scaled by the stellar escape speed $v_{\rm e}$,
densities are in $\rho_{c\ast} \equiv \Mdot_{B=0}/4\pi \Rstar^2 v_{\rm e} $,
and 
temperatures are in $T_\infty = (3/16) \mubar \vinf^2/k$.

The global parameters are: 
the maximum closed loop apex $R_c$, fixed here to $R_c/\Rstar = 4$;
the ratio of wind terminal speed to surface escape speed, fixed here to $\vinf/v_{\rm e} =3$;
the loop apex smoothing length $\hzd$, with standard value $\hzd/\Rstar = 0.3$;
and the cooling parameter $\chi_\infty$ (defined in equation \ref{eq:chiinf}).
All  components also have floor temperature set to the stellar  effective temperature,
taken here to be $T_{\rm eff} = 30,000$\,K.
 
Finally, for all three components, the flow is along the dipole field line, with thus vectorial direction given by equation (\ref{eq:bhatrmu}), with a radially positive sense for the upflow and post-shock components, and a negative sense for the cooled downflow.

\section{Comparison to MHD simulations}  
\label{sec:MHD}

To assess the potential for these simplified ADM scalings to provide a basis for deriving observational diagnostics, 
let us next compare their predictions to time-averaged results from full MHD simulations.
 
For consistency with the sample plots given above, we again choose as a standard the ADM model with a maximum loop closure radius $R_c = 4 \Rstar$.
For corresponding MHD simulations, we use the standard stellar parameters of paper I, but now with a polar magnetic field $B_p=5000$\,G, which gives an Alfv\'{e}n radius $R_A \approx  R_c$.
The time averaging begins at $t=500$\,ks, when the structure has fully relaxed from the initial condition, and extends to $t=4400$\,ks, representing about ten cycles of mass build-up and dynamical infall. 
To ensure the north-south symmetry of an arbitrarily long time average (or from azimuthal averaging of a full 3D model; see figure 3 of \citet{Uddoula13}),  we also carry out a north-south averaging of all the time-averaged quantities from this 2D MHD simulation.

The remainder of this section focuses on the relatively cool ($T \sim T_{\rm eff} \sim 10^4$\,K) 
material in the outflowing wind and the cooled downflow.
This is then used in \S \ref{sec:Balmer} to derive signatures in
recombination-based emission lines like H$\alpha$.
Section \ref{sec:xrays}  next focuses on X-ray emission from the hot post-shock gas, and its absorption from the wind and cooled downflow.

\subsection{Cooled downflow}
\label{sec:adm-mhd-cd}

First, for closed loop lines with apex radii $r_m \le R_c = 4 \Rstar$, the top row of figure \ref{fig:adm-vs-mhd} plots ADM scalings for the density (with apex smoothing length $\hzd=0.3 \Rstar$; left panel), and the latitudinal and radial components of velocity (middle and right panels) in the cooled downflow component.
The lower panels compare corresponding time-averaged quantities from the MHD simulations.
For simplicity, the lower left shows the time averaged density $\left < \rho \right >$, but 
because emission profiles depend on the velocity of material weighted by its density-squared emission measure, the middle and right panels show the time averages of the speeds weighted by the density squared.

Note that in the MHD model the signatures of trapped material in closed magnetic loops is radially more extended than in the idealized, dipole form of closed loops in the ADM model, due to the dynamical 
interaction of the field and radiatively driven wind outflow.
The dynamical variation of  infall episodes means the density is less equatorially concentrated in the MHD vs. ADM models, but overall the MHD model density, as well as the radial and latitudinal velocities, show a clear boundary change linked to closed loop geometry, much as in the ADM case.

To make these comparisons semi-quantitive, the assumed stellar parameters of the  MHD simulation 
(which are the same as the standard model of paper I)
imply a surface escape speed $v_{\rm e} \approx 700$\,km\,s$^{-1}$, and 
 a characteristic wind-fed loop density $\rho_{c\ast} \approx 1.5 \times 10^{-13}$\,g\,cm$^{-3}$.
Using these to scale the quoted CGS values in the colorbars for the MHD simulation in the lower row of figure \ref{fig:adm-vs-mhd}, we see that the color levels are in quite good agreement with the ADM case.

\begin{figure}
\begin{center}
\vfill
\includegraphics[scale=0.65]{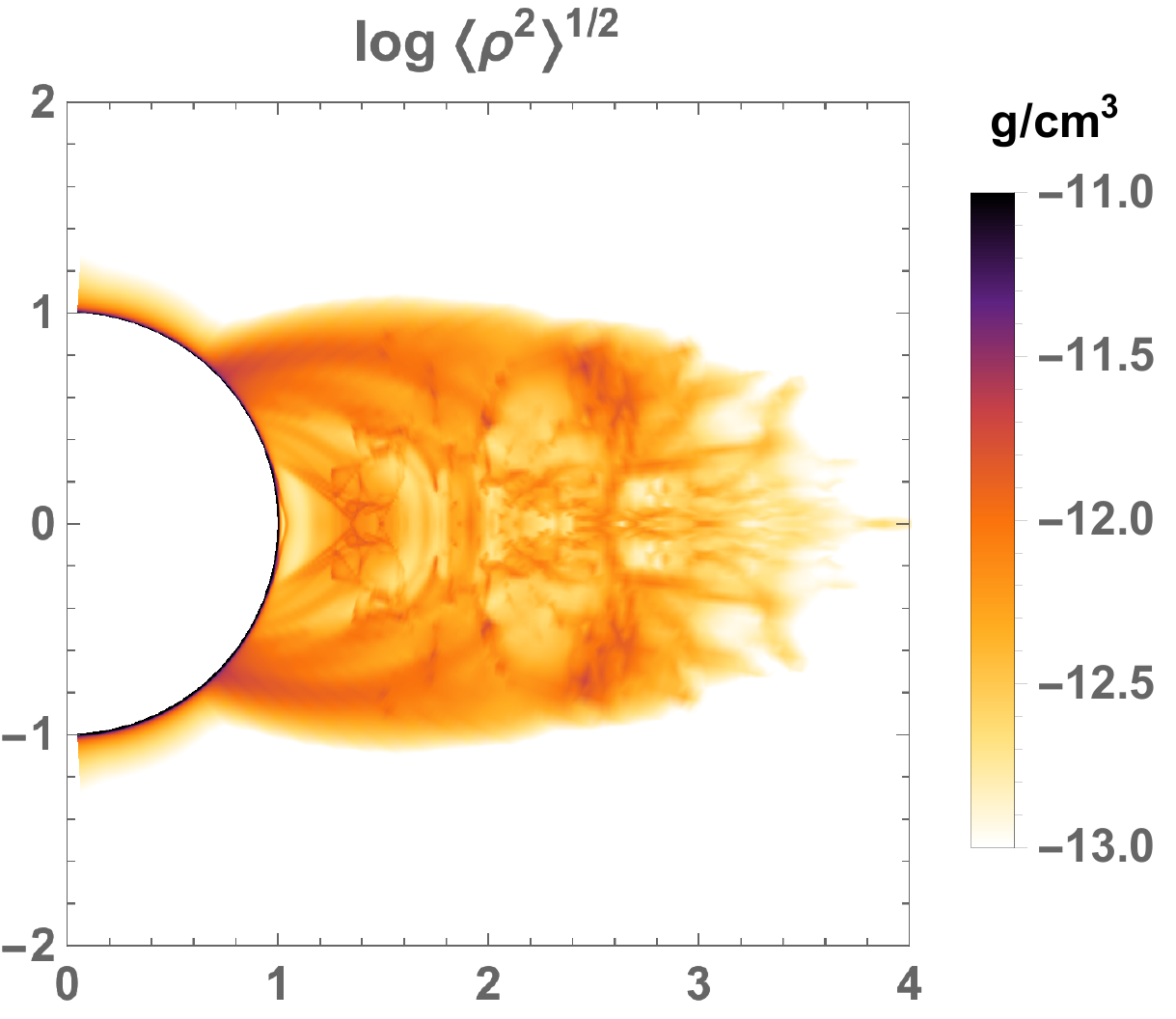}
\caption
{
For the MHD model shown in figure \ref{fig:adm-vs-mhd}, spatial variation of time-averaged 
rms density $\left < \rho^2 \right >^{1/2}$.
Note that the overall form is very similar to the time-averaged density plot in  figure \ref{fig:adm-vs-mhd}d, but the colorbar scale here is enhanced by about a factor 7, due to the clumping associated with the complex infall of compressed material.
}
\label{fig:rhorms}
\end{center}
\end{figure}

However, one key aspect of the cooled downflow not accounted for in the ADM steady-state picture is that in full MHD simulations, the material infall actually occurs in sporadic intervals of highly compressed, localized streams. 
This implies a significant enhancement in the mean of the density-{\em squared}, which for the two-body collision or recombination processes that underlie line emission is what sets the overall emission strength.
For this  MHD model, figure~\ref{fig:rhorms} plots  now the time-averaged {\em rms} density $ \left < \rho^2 \right >^{1/2}$.
While the overall form is very similar to the  density plot in figure~\ref{fig:adm-vs-mhd}d, note that, due to the clumped infall, the colorbar scale is now enhanced by about a factor 7.
The associated ``clumping factor'' $f_{cl} \equiv \left < \rho^2 \right >/\left < \rho \right >^2$,
which effectively sets the level of  emission enhancement relative to a smooth model with the same average density, is thus typically increased by several factors of ten.
In applying the ADM scalings to model such optical emission lines, one should account for this clumping by enhancing the emission by a clumping factor $f_{cl}$ of this order.
The next section provides a first example for the case of hydrogen H-$\alpha$ line emission.

\section{Application to Observational Diagnostics: Optical Line Emission}
\label{sec:Balmer}

With this background, let us now derive scaling relations and perform a first diagnostic application of the ADM model for the hydrogen H-$\alpha$ line.
A key advantage with H$\alpha$ is that under typical O-star wind conditions its atomic level populations remain quite close local thermodynamic equilibrium (LTE) with respect to the \textit{real} population of ionized hydrogen  \citep[e.g.,][]{Puls96, Sundqvist11}, allowing one to use relatively simple methods to analyze this line also in magnetic O-stars \citep{Sundqvist12c}.  

\subsection{ADM-modified scaling relation for H$\alpha$ emission} 
\label{sec:admha}

The principal scaling of H$\alpha$ emission in OB-star winds comes from considering the line optical depth $\tau$ in the Sobolev approximation. For Doppler width $\Delta \nu \equiv v_{\rm th} \nu_0/c$ 
and directional Sobolev length $L = v_{\rm th}/(dv_{\rm n}/dn)$, this is   
\begin{equation}
  \tau = A \rho^2 L / \Delta \nu,  
  \label{Eq:tau_sob} 
\end{equation} 
where we have absorbed atomic constants of the H$\alpha$ transition and  dependencies on electron temperature and occupation number 
densities into the parameter $A$, as given in Appendix C \citep[see also][]{Puls96, Petrenz96}. 
For an observer viewing from above the magnetic pole, equation (\ref{Eq:tau_sob}) can be used to derive an optically thin emission-measure scaling law for the ADM model (see Appendix C) for a polar-view observer: 
\begin{equation} 
	W_{\rm ADM} \sim \frac{\dot{M}_{B=0}^2 \,f_{\rm cl}}{R_\star^3 v_{\rm e}^2} \frac{f(R_c)}{\hzd/R_\star}, 
	\label{Eq:ewha} 
\end{equation}      
where the function $f(R_c)$ describes the dependence on the size of the closed-loop magnetosphere. 
Equation (\ref{Eq:ewha}) illustrates explicitly how the standard scaling for non-magnetic wind emission is modified here by the two magnetic parameters $\hzd$ and $R_c \approx R_{\rm A}$ (essentially setting the ADM disc-thickness and size), and comparison to full radiative transfer calculations (Sect. 4.2) indicates that this  simple scaling law captures quite well the principal scaling of ADM H$\alpha$ emission under typical OB-star conditions. 

\subsection{A first diagnostic application} 
\label{sec:firstdiag}

\begin{figure*}
\begin{minipage}{8.5cm} 
 \resizebox{\hsize}{!}
    	        {\includegraphics[angle=90]{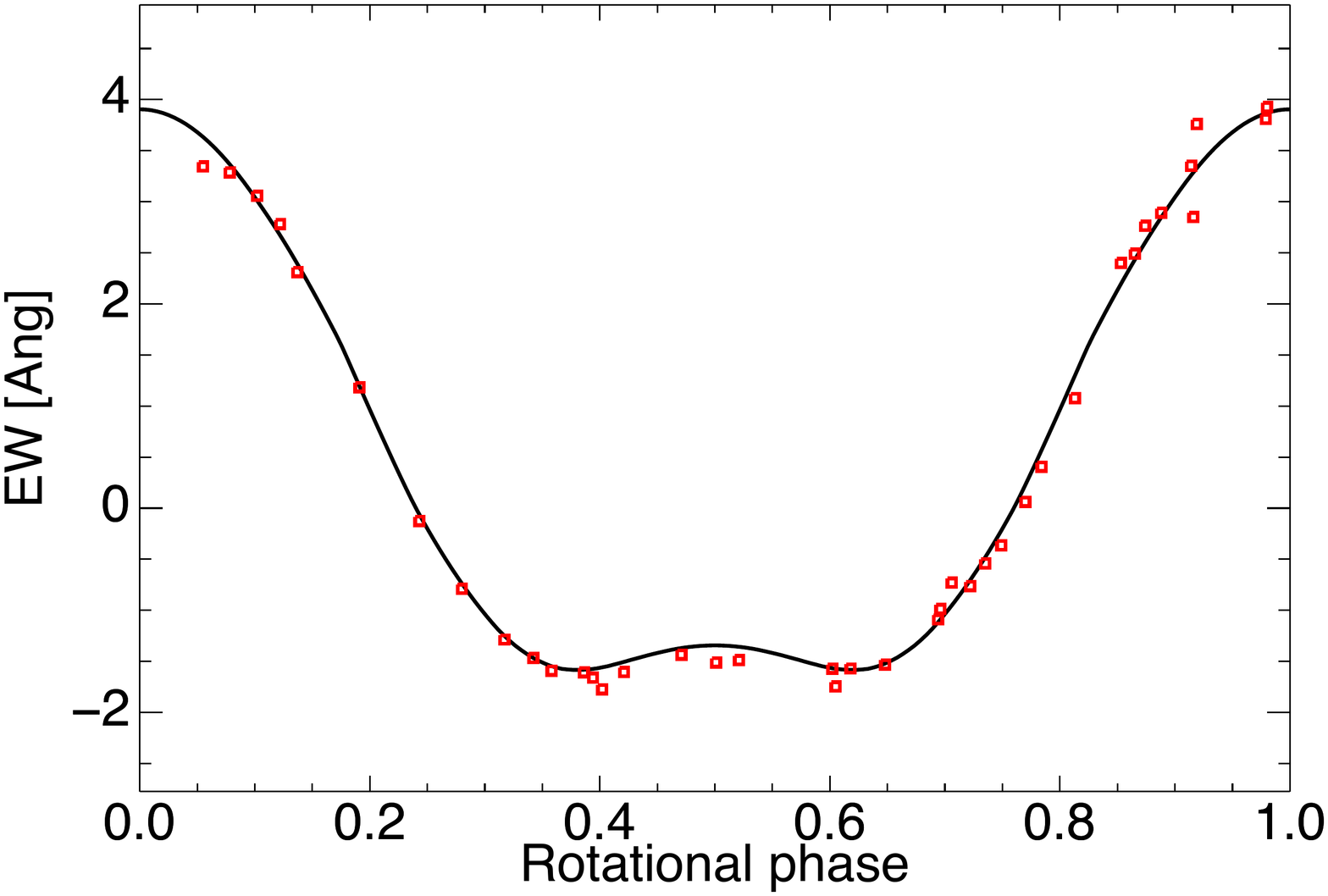}}	 
		\centering
\end{minipage} 
\begin{minipage}{8.5cm} 
 \resizebox{\hsize}{!}
              {\includegraphics[angle=90]{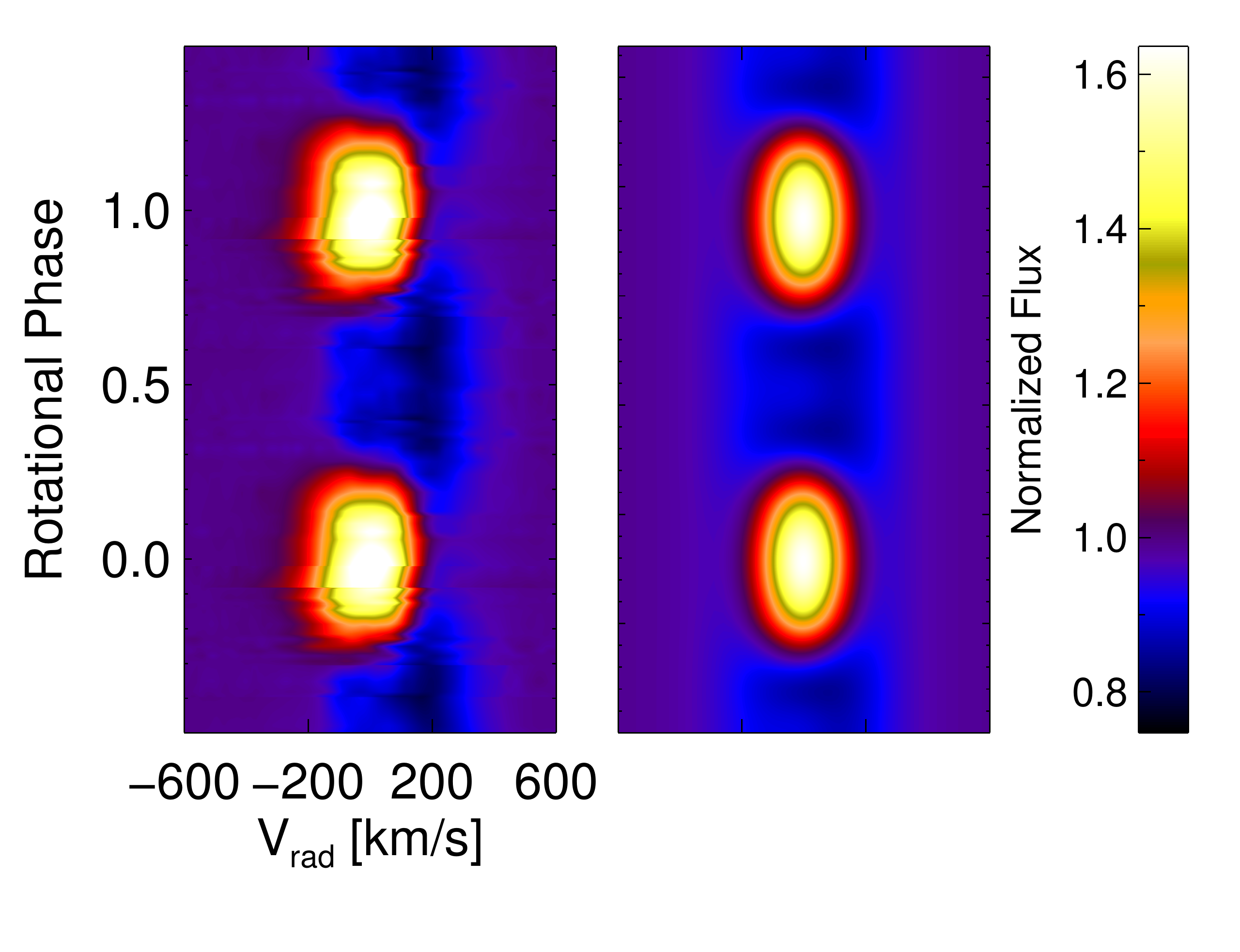}}	 
		\centering
\end{minipage} 
\caption{
\textbf{Left:} Fit to observed H$\alpha$ equivalent width (EW) light-curve 
(EW in units of Angstrom, with net emission counted positive)
of HD191612 (red squares) as function of rotational phase, using the 
ADM model (black solid line). \textbf{Right:} Normalized flux dynamic spectra of 
observations (left panel) and simulations (right panel).
}
\label{Fig:ha}
\end{figure*}

Building on the simple scaling analysis above, let us now make a first application toward using optical emission lines like H$\alpha$ to diagnose the winds and magnetospheres of slowly rotating magnetic OB-stars. 
To compute synthetic spectra from the steady-state ADM density and velocity structure, we follow \citet{Sundqvist12c} and solve the formal solution of radiative transfer in a 3D cylindrical system for an observer viewing from angle $\alpha$ with respect to the magnetic axis. 
Defining $\beta$ and $i$ as the angles that the magnetic axis and the observer line-of-sight make with the rotation axis,
the variation of $\alpha$ with rotational phase $\Phi$ is given by
\begin{equation} 
	\cos \alpha = \sin \beta \cos \Phi \sin i + \cos \beta \cos i
	\, .
\end{equation} 
We solve the transfer equation only in the infall compoent, assuming H$\alpha$ departure coefficients and an electron temperature structure calibrated by 1D NLTE model atmosphere  calculations (e.g., \citealt{Puls96}; \citealt{Sundqvist11}), and using an input photospheric H$\alpha$ line-profile as lower boundary condition (see \citealt{Sundqvist12c}, for more details).

The rotational phase variation of the emission can now be used to derive 
the magnetic geometry of the oblique rotator, and the absolute level of H$\alpha$ emission 
further constrains the rate $\dot{M}_{\rm B=0}$ by which the magnetosphere is fed by 
radiatively driven wind plasma \citep[see also][]{Wade15}; this latter diagnostic is 
directly evident through the scaling in 
equation (\ref{Eq:ewha}), and is analogous to how H$\alpha$ emission 
from non-magnetic O-stars constrain the wind mass-loss rate. Moreover, since the smoothing
length parameter $\hzd$ affects the computation of line optical 
depth quite differently depending on viewing 
angle, the level of emission contrast between maximum 
and minimum phases ('high' and 'low' states) can, in principle, also be used to 
constrain this ADM parameter.

As an explicit example of the potential diagnostic power of 
our model, Figure~\ref{Fig:ha} shows quite remarkably good fits to 
the observed H$\alpha$ light-curve and dynamic spectra \citep{Howarth07} 
of the slowly rotating magnetic O-star HD191612 \citep{Wade11c, Sundqvist12c}.
We use the stellar parameters derived for HD191612 by \citet{Howarth07}  
(effective escape speed $v_{\rm e} = 600$\,km/s and stellar 
radius $R_\star = 14.5 R_\odot$), and a maximum 
loop-closure radius set additionally by the observed dipole magnetic field strength 
\citep{Howarth07, Wade11c}, $R_c \approx r_{\rm A} \approx \eta_\star^{1/4} R_\star \approx 
3.5 R_\star$ for wind confinement-parameter
$\eta_\star \equiv B_\star^2 R_\star^2/(\dot{M}_{\rm B=0} v_\infty)$  \citep{Uddoula02}.  
Since $R_{\rm C}$ depends only weakly on mass loss and wind terminal speed 
(to the 1/4 power) and all other parameters here are known from 
observations, we may safely keep the closure radius fixed during the fitting.

Matching the level and variability contrast of the 
observed H$\alpha$ line profiles yields here $\hzd \approx 0.5 R_\star $ 
and $\dot{M}_{\rm B=0} \sqrt{f_{\rm cl}} \approx 5 \times 10^{-6} \, M_\odot \, yr^{-1}$. 
This is approximately a factor of three higher than the mass-loss rate 
used by \citet{Sundqvist12c} to model the H$\alpha$ rotational phase variation 
in HD191612 directly from MHD simulations, and reflects the fact 
that the steady-state ADM model does not account for density-squared 
enhancements produced by the highly clumped streams of infalling material
(see discussion above)\footnote{While the modeling in 
\citet{Sundqvist12c} naturally includes this effect, also this analysis 
neglects the stochastic, small-scale inhomogeneities caused by the instability 
inherent to line-driven winds (Owocki, Castor \& Rybicki 1988); simulating 
also this strong instability requires a non-local treatment of the radiation line 
force and has yet to be implemented within any MHD model.}. 
\S3.1. suggests clumping factors of several factors of ten, and 
adopting here for example $f_{\rm cl} \approx 50$ would then 
imply $\dot{M}_{\rm B=0} \approx 0.7 \times 10^{-6} M_\odot \, yr^{-1}$, 
which is in quite good agreement with predictions from radiatively 
driven wind theory \citep{Vink00}.  
Using these parameters, the magnetic geometry is then also derived from the 
rotational phase variation and
shape of the H$\alpha$ equivalent-width curve, resulting here in a degenerate couple $(\beta, i) = (i, \beta) \approx 23, 73^{\circ}$, which agrees well with the  magnetic geometry derived from spectropolarimetry \citep{Wade11c}. 

In order to match the observed line widths, we have further convolved the dynamic synthetic ADM spectra presented in  Figure~\ref{Fig:ha} by a 150 km/s isotropic Gaussian 'macro-turbulence'. 
While it is not surprising the steady-state ADM models show too little velocity dispersion, we note that such extra broadening is actually required also when modeling H$\alpha$ directly from MHD  simulations \citep{Sundqvist12c, Uddoula13}.

While the analysis here shows very good fits of 
the ADM model to the hydrogen H-$\alpha$ line in HD191612, the key aim of this first study has been to demonstrate the diagnostic potential of the model, rather than to obtain perfect 
estimates of all stellar, wind, and magnetic parameters. As discussed 
further in Appendix C, future detailed parameter-studies of different regimes 
will be required to fully evaluate, e.g., the accuracy of ADM-derived $\dot{M}_{\rm B=0}$.
 
\section{Application to Observational Diagnostics: X-rays}
\label{sec:xrays}

\subsection{Hot post-shock gas and X-ray emission}
\label{sec:hpsg-xrays}

The results in section \ref{sec:hpsg} for the temperature and density of the hot post-shock gas provide a basis for computing the X-ray emission from such DM's.
Following Appendix B of paper I, for a given gas temperature $T$ let us write the spectrally integrated (mass-weighted) emission function above a specified X-ray energy $E_x$ as 
\beq
{\bar \Lambda}_m (T,E_x) \approx \Lambda_m e^{-E_x /kT}
\, ,
\label{eq:lammTEx}
\eeq
where the approximation expresses this in terms of the total (mass-scaled) cooling function $\Lambda_m$ and a simple ``Boltzmann'' factor in the ratio $E_x/kT$.
In the context of the present ADM model, let us characterize the threshold energy $E_x$ in terms of its ratio
$\epsilon_x \equiv E_x/kT_\infty $ to the maximum shock energy $kT_\infty$. 
The associated spectrally integrated volume emissivity in the hot, post-shock region is then given by
\beq
\eta_x (r,\mu, \epsilon_x) =  \
\Lambda_m \left ( \rho_h (r,\mu) \right )^2 \, e^{- \epsilon_x T_\infty/T_h(r,\mu)}
\, .
\label{eq:etah}
\eeq
For the full MHD simulations of paper I, an analogous simplified form (\ref{eq:etah}) was found to reproduce quite well the spatial distribution of X-ray emission computed from a full integration of the atomic emissivity above the given threshold energy $E_x$.
Appendices A and B here provide a further analysis of how the ADM model can be used to derive both the differential emission measure (DEM), as well as a shock-temperature distribution $p(T_s)$.

\subsection{X-ray emission in ADM vs.\ MHD simulations}
\label{sec:xray-adm-vs-mhd}

\begin{figure*}
\begin{center}
\includegraphics[scale=0.43]{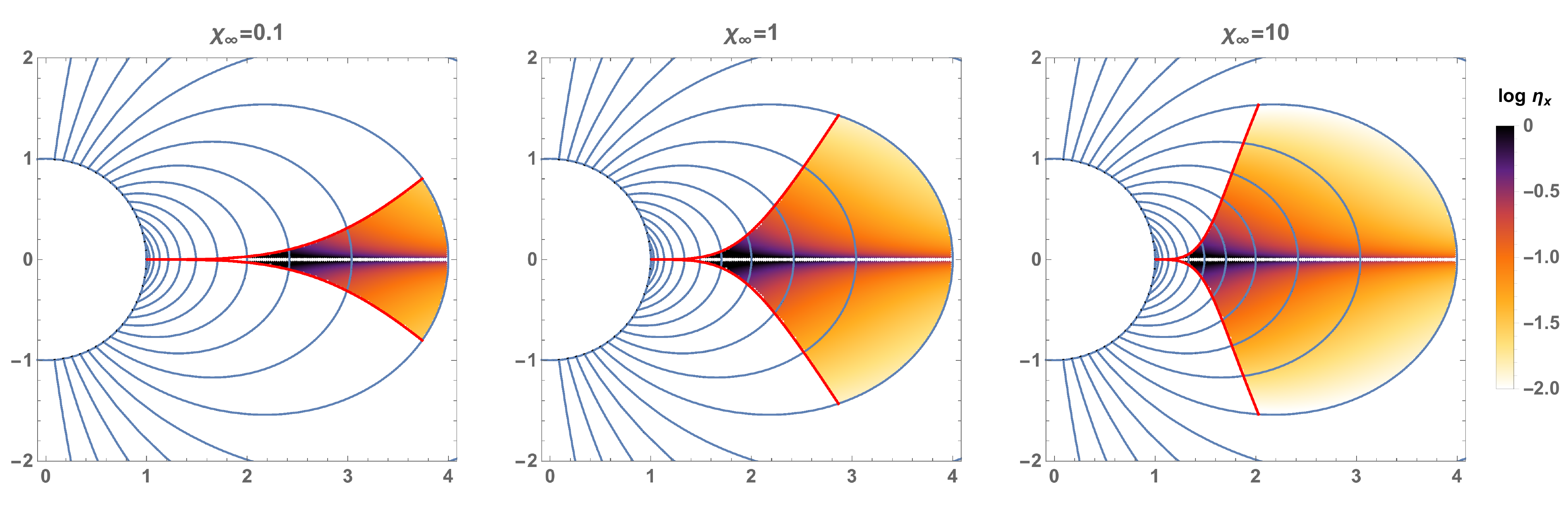}
\\
\includegraphics[scale=0.47]{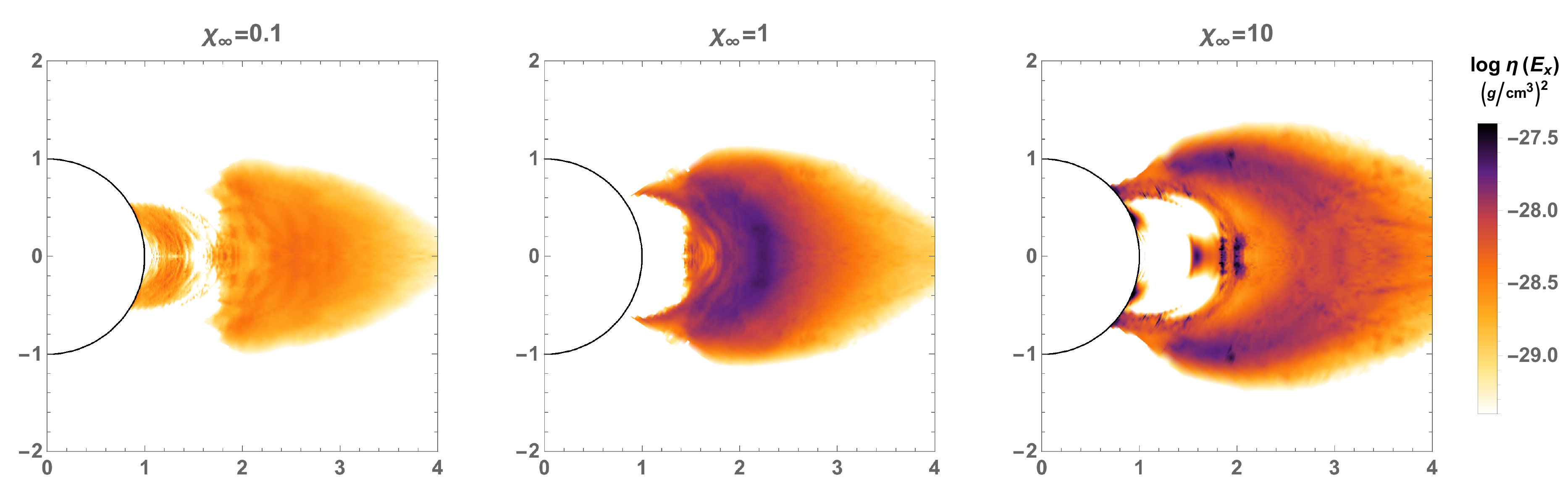}
\caption
{Top: For X-rays above a scaled energy $\epsilon_x = E_x/kT_\infty=0.017$, log of ADM emissivity $\eta_x $ 
(scaled by $\Lambda_m \rho_{w\ast}^2 \, \chi_\infty$) for $\chi_\infty=$ 0.1, 1, and 10  (left, middle, right).
Bottom:  For  MHD simulations with the parameters of the standard model in paper I,  log of the time-averaged X-ray emissivity above threshold energy $E_x = 0.13$\,keV, for cooling efficiencies tuned to give $\chi_\infty = 0.1$, 1, and 10;
the middle ($\chi_\infty = 1$) case is the same simulation used for figure~\ref{fig:adm-vs-mhd}.
For the terminal speed $v_\infty = 2500$\,km/s of the associated non-magnetic wind, the  terminal shock energy $kT_\infty =  7.5$\,keV implies the same threshold energy fraction $\epsilon_x \equiv E_x/kT_\infty = 0.017$ used in the corresponding ADM models.
As in figure~\ref{fig:adm-vs-mhd}, the MHD simulation data here has been north-south symmetrized to provide clearer comparison with the symmetric ADM model.
}
\label{fig:adm-vs-mhd-etah}
\end{center}
\end{figure*}

Let us next make a direct comparison of the spatial distribution of the X-ray emission in ADM model vs.\ that found in MHD simulations.
For  the same 3 cooling parameter values ($\chi_\infty =$ 0.1, 1, and 10) used for the hot-gas temperature and density plots in 
figure~\ref{fig:hall}, the top row of figure 
\ref{fig:adm-vs-mhd-etah} plots the associated variation of the X-ray emissivity $\eta_x$, scaled here by $\Lambda_m \rho_{w\ast}^2 \chi_\infty$.
The bottom row compares results for the MHD simulations, showing now the time-averaged Boltzmann corrected emission (divided by $\Lambda_m$, to give results in units of a density squared) for gas above a threshold temperature of $T_x = 1.5$\,MK, corresponding to an X-ray threshold energy  $E_x = kT_x \approx$\, 0.13\,keV.
For the terminal speed $v_\infty = 2500$\,km/s of the associated non-magnetic wind, the  terminal shock energy $kT_\infty =  7.5$\,keV then implies a threshold-energy ratio $\epsilon_x \equiv E_x/kT_\infty = 0.017$, which is thus the value used in the corresponding ADM models.
As in figure~\ref{fig:adm-vs-mhd}, the MHD simulation output here has been north-south symmetrized to provide clearer comparison with the symmetric ADM model.

Paper I showed that the volume integrated X-emission from MHD simulations can\footnote{with a factor 0.2 reduction associated with the duty cycle of X-ray emission intervals between infall events; see paper I.} 
be well modeled with an ``XADM'' analysis that is grounded in the same basic ADM scalings used here
(see eqns.\ \ref{eq:dlxdmus1}-\ref{eq:dlxdmusf} of Appendix A).
However, figure~\ref{fig:adm-vs-mhd-etah} shows that, while the idealized ADM model predicts a marked concentration of the X-ray emission in the equatorial region around the loop tops, the time-averaged X-ray emission in the MHD simulations is much more spatially extended about the magnetic equator.

Indeed one can identify two distinct  X-ray emitting regions, with distinct physical origins.
The X-rays from inner loops arise from sporadic intervals of ``siphon'' flow between loop footpoints.
This effect is still poorly understood  \citep{Bard15}, but it makes only a minor overall contribution to the total X-ray emission, and is not included in the ADM model.

The second, outer-loop component arises more directly from the collisional shock and retreat that is characterized by the hot post-shock gas in the ADM analysis.
The time-variable `sloshing' of hot post-shock gas in the MHD simulations makes its associated time-averaged X-ray emission much more extended, but its radial onset in the MHD simulations corresponds quite well with the inner edge of the shock retreat in the ADM model.

The upshot is that, while the ADM model exaggerates the equatorial concentration of the latitudinal distribution of X-ray emission, it provides a good general description of both its overall spatially integrated value (paper I), and it radial distribution and extension away from the stellar surface.

\subsection{X-ray absorption}
\label{sec:xray-abs}

\begin{figure*}
\begin{center}
\includegraphics[scale=0.61]{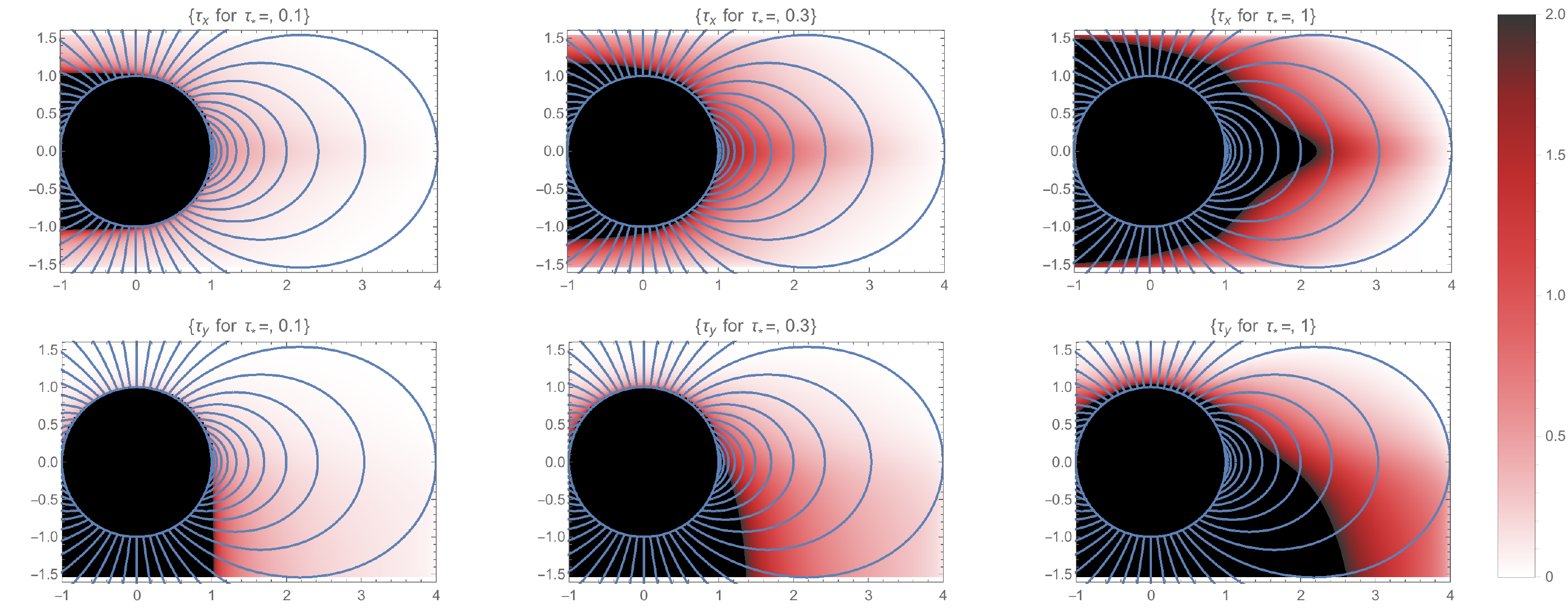}
\caption{
Spatial variation of optical depth for bound-free absorption of X-ray emission by both the cool downflow and wind outflow components of the ADM model, as well as by occultation of the opaque star.
The top row shows results for a distant observer to the right, with an equator-on view, while
the bottom row is for an observer at the top, with a pole-on view.
The model assumes an apex smoothing length $\hzd=0.1 \Rstar$, and a terminal speed $\vinf = 3 v_{\rm e}$ for 
a corresponding unmagnetized wind.
The left, middle and right columns show cases with a corresponding wind 
optical depth $\tau_\ast  \equiv \kappa {\dot M}/(4 \pi \Rstar \vinf  )= $ 0.1, 0.3 and 1.
}
\label{fig:tauxyvec}
\end{center}
\end{figure*}

While the shock-heated X-ray emitting plasma is expected to be mostly optically thin to its own radiation, the cool components of the wind and magnetosphere may very well be optically thick. The continuum opacity in this modestly ionized plasma is due primarily to inner-shell photoionization (bound-free opacity) of metals, and is wavelength-dependent, with generally higher opacities at longer wavelengths. The associated attenuation of the X-rays may lead to potentially observable effects that should have diagnostic power, especially in phase-dependent spectra.

In single O stars with embedded wind shock X-rays, signatures of wind absorption are seen in stars with higher mass-loss rates 
($\Mdot \gtrsim {\rm few} \times 10^{-7}  \Msun$yr$^{-1}$) via the distortion of X-ray emission line shapes \citep{Cohen14}.
 Broadband signatures of X-ray absorption, hardening the emergent spectrum, are also seen in both high-resolution grating spectra and lower resolution CCD spectra of single O stars 
 \citep{Leutenegger10, Cohen11}.
Similar X-ray absorption by the fast wind in the open field regions of magnetic O stars should be expected, while the contribution from the magnetosphere, with its steady-state upflow, shock, and downflow cycle, should be comparable to that from a spherical unmagnetized wind. In fact, because of the slower velocity and higher density of the confined wind in the magnetosphere, the degree of X-ray attenuation in the cooled downflow component of the ADM is potentially large. 

The bound-free X-ray opacity in the wind and cool magnetospheric plasma is expected to more or less monotonically increase with wavelength through most of the X-ray bandpass, as shown in, e.g., figure~2 of \citet{Cohen14}.
This is because for each ion that contributes to the bound-free opacity, the opacity is strongest closest to the threshold set by the ionization potential and decreases strongly toward higher energies. The opacity at any wavelength is the sum of the contributions from all abundant ions. 
The biggest uncertainty and potential cause of star-to-star variation in the X-ray opacity is likely due to helium, which may be singly ionized in some cases -- and would thus contribute significant opacity at longer wavelengths -- and may be fully ionized -- and thus contribute no bound-free opacity -- in other cases. 
The size of this helium ionization effect can be seen in figure~3 of \citet{Cohen14}.
Note that the wind opacity at a fiducial photon energy of 1 keV (12 \AA) corresponds to a cross section per hydrogen atom of roughly $10^{-22}$ cm$^{2}$. 

We can characterize the overall optical depth of a given ADM model by scaling the column density to the quantity $\tau_{\ast,\lambda} = \kappa_\lambda \Mdot_{B=0}/4\pi\Rstar\vinf$. 
Thus a given mass-loss rate and a given opacity at a particular wavelength corresponds to a particular $\tau_{\ast,\lambda}$ value, and the optical depth is degenerate in these two parameters, being proportional to their product. 
In figure~\ref{fig:tauxyvec} we show optical depth maps for three different $\tau_\ast$ values, each computed for two different viewing geometries: one in the magnetic equator and one over the magnetic pole. 
For the largest characteristic optical depth value shown, $\tau_{\ast,\lambda}$=1, which is expected for longer observed wavelengths in a star with a theoretical, non-magnetic mass-loss rate of less than $10^{-6} \Msun$yr$^{-1}$, there is significant magnetospheric absorption of X-rays in even the front hemisphere in both the edge-on and pole-on views. 
This is in addition to occultation by the star itself, which will only be relevant for the edge-on view if the X-ray emitting plasma is in the magnetic equator. 
The figure shows our standard model with $R_c = 4 \Rstar$. A larger closed magnetosphere should produce even more attenuation. Although there will be some variation depending on the assumed location of the X-ray emitting plasma. 
As discussed in the previous section, the ADM models show a concentration of the weighted X-ray emissivity in the magnetic equatorial plane and near $2 \Rstar$, with very little dependence on $\chi$, while the MHD simulations show a somewhat more complex situation (Figure~8). 
 
From an observational point of view, no significant X-ray absorption is seen in the phase-resolved Chandra grating observations of the prototype magnetic O star, $\theta^1$Ori\,C, which has $R_A \sim 2 \Rstar$ \citep{Gagne05}.
Numerical MHD simulations show magnetospheric column densities of order $10^{21}$ cm$^{-2}$, consistent with negligible attenuation 
\citep[][figure 10]{Uddoula13,Petit15}.
On the other hand, significant X-ray absorption is detected in the low-resolution Chandra spectra of NGC\,1624-2 
\citep{Petit15},
which is the O star with the strongest magnetic field and the largest magnetosphere ($R_A \approx 11$ vs.\ $\approx 2 \Rstar$ for $\theta^1$Ori\,C).  More X-ray absorption is seen in NGC\,1624-2 when it is observed edge-on than when it is observed at a nearly pole-on phase, a viewing angle modulation that the ADM model is well suited to model. Such a large magnetosphere, with its very strong surface field, is prohibitively expensive to model using numerical MHD. 

\section{Summary and Future Outlook}
\label{sec:conclusions}

The ADM analysis here provides readily computable formulae for the basic hydrodynamic quantities -- density, temperature, and velocity -- for each of the three components of a wind-fed dynamical magnetosphere -- wind upflow, hot post-shock gas, and cooled downflow.
Comparison with time-averaged values derived from detailed MHD simulations show, with some caveats, quite good general agreement.
As such, this  ADM formalism can provide a conceptually and computationally much simpler basis for synthesizing observational diagnostics, and for deriving broad scaling relations for how these depend on stellar, wind, and magnetic parameters.

For X-ray spectral bands, \S \ref{sec:xray-adm-vs-mhd} compares directly the X-ray emission in ADM vs.\ MHD models,
while \S \ref{sec:xray-abs} discusses how observed X-ray spectra could be affected by absorption from the cool components.
Appendices A and B present a general scaling analysis of how the ADM model can be used to derive both the differential emission measure, as well as a shock-temperature distribution $p(T_s)$.
This augments the XADM analysis of paper I, and so builds on the promising  agreement of the derived scaling laws with observations \citep{Uddoula14,Naze14}.

To illustrate ADM spectral synthesis in optical emission lines, we exploit the relative simplicity of the H$\alpha$ line formation process in O-type stars. 
\S~\ref{sec:Balmer} explicitly demonstrates the potential diagnostic power of the ADM model by a first successful application to observations 
of the rotational phase variation of H$\alpha$ emission from the O-star HD191612.
The remarkably good agreement provides constraints on key physical parameters like magnetic geometry and overall mass loss rate $\dot{M}_{\rm B=0}$, thus illustrating the utility of the ADM formalism even for modeling individual stars with DM's.
The analysis in Appendix C provides also general scaling relations with stellar, wind, and magnetospheric parameters.

But the simple, steady-state nature of the ADM model paves the way for future applications that require more elaborate NLTE radiative transfer. 
For example, in magnetic O-stars optical helium lines like HeII 4686${\rm \AA}$ show clear signatures of being formed in a DM \citep[][]{Grunhut12, Wade15}, and recent observations of magnetic massive stars in the infra-red \citep[e.g.,][]{Oksala15} suggest a strong influence of the magnetosphere also for key diagnostics in that waveband. 
Moreover, the physical explanation of the so-called Of?p morphological phenomena \citep{Walborn72} of magnetic O-stars is very likely 
related to a complex formation scenario of nitrogen and carbon spectral lines in a DM. 

In summary, much as complex computer codes like {\sc cmfgen} \citep{Hillier98} or {\sc fastwind} \citep{Puls05} nowadays are routinely applied for spectroscopic analyses of non-magnetic hot stars with winds, we envision that the ADM model presented here lays the groundwork for development of NLTE radiative transfer tools for detailed spectroscopic analysis of magnetic massive stars. 

\section*{Acknowledgments}

This work was supported in part by SAO Chandra grant TM3-14001A 
and NASA  Astrophysics Theory Program grant NNX11AC40G, awarded to the University of Delaware. 
AuD acknowledges support by NASA through Chandra Award number TM4-15001A and 16200111 issued by the Chandra X-ray Observatory Center which is operated by the Smithsonian Astrophysical Observatory for and behalf of NASA under contract NAS8- 03060.
AuD also acknowledges support for Program number HST-GO-13629.008-A provided by NASA through a grant from the Space Telescope Science Institute, which is operated by the Association of Universities for Research in Astronomy, Incorporated, under NASA contract NAS5-26555.
JOS acknowledges funding from the European UnionÕs Horizon 2020 research 
and innovation programme under the Marie Sklodowska-Curie grant agreement No 656725.
DHC acknowledges support of  Chandra grant TM4-15001B to Swarthmore College.
RHDT acknowledges NASA ATP Grant NNX08AT36H to the University of Wisconsin. 
VP acknowledges partial support of NNX15AG33G from NASA's XMM-Newton Guest Observer Facility, 
and NASA Chandra grants TM4-15001C and GO3?14017A.
We thank D.\ Kee and G.\ Wade for many fruitful discussions. 
We thank the anonymous referee for many constructive criticisms and comments that helped improve the final version of this paper.
  
\bibliographystyle{mn2e}
\bibliography{OwockiS}

\phantom{.}

\appendix

\section{Cooling of post-shock material along fixed loop line.}

For plasma of density $\rho$ at temperature $T$, we can approximate the local volume emissivity of X-rays above a given energy $E_x$ as,
\beq
\eta (E_x) = \rho^2 \frac{{\bar \Lambda}(T,E_x)}{\mu_e \mu_p} \approx \rho^2 \Lambda_m (T) e^{-E_x/kT} \,  .
\label{eq:etaxdef}
\eeq
Here ${\bar \Lambda}(T,E_x)$ is the integrated plasma emission function above energy $E_x$, 
plotted in figure A2 of paper I; 
as illustratred by the corresponding dashed curves in that figure, the latter approximation expresses this in terms of the total plasma emission, $\Lambda_m (T) \equiv {\bar \Lambda}(0, T)/\mu_e \mu_p$, times a Boltzmann factor.

Let us next integrate this over a volume trace along a fixed, closed magnetic loop line with surface footpoint at a colatitude set by $\mu_\ast$, and with a surface field-line-projection $\mu_B$. In terms of the associated local area $A \sim 1/|B|$ of the magnetic flux tube along the field line coordinate $b$,  the contribution to X-ray luminosity per colatitude interval $d\mu_\ast$ is
\beqa
\frac{dL_x }{d\mu_\ast} (\mu_\ast,E_x) 
&\approx& \mu_B \int_{b_s}^{b_m}   \Lambda_m  \rho^2 \, e^{-E_x/kT} A \, db
\label{eq:dlxdmus1}
\\
&=&
 - \frac{5k}{2{\bar \mu}} 
\mu_B  \int_{b_s}^{b_m} 
  \rho v A \, \frac{dT}{db}  e^{-E_x/kT}  \, db
 \\
&=&
 \frac{5kT_s}{2{\bar \mu}} 
\mu_B^2  \Mdot_{B=0} \, \int_0^{T_s} e^{-E_x/kT}  \, \frac{dT}{T_s} 
\\
&=&
 \frac{15}{16} \,
\mu_B^2  \frac{\Mdot_{B=0} \vinf^2}{2} \, w_s^2 f_x (T_s,E_x)
\ .
\label{eq:dlxdmusf}
\eeqa
The second and third forms use the energy equation (\ref{eq:dTdb1}) and the field line mass flux equation (\ref{eq:mds}), while the last equality uses the fraction $f_x (T_s,E_x)$  of shock energy emitted above the threshold $E_x$, as given by eqns.\ (35)-(37) of paper I.
This result recovers the basic XADM scaling derived in section 5 of paper I; integration over colatitude gives equation (39) there\footnote{
The factor $15/16$ stems from the isobaric approximation used in the energy equation (\ref{eq:dTdb1}) to model the post-shock cooling, since this ignores the post-shock kinetic energy component, which is $(1/4)^2=1/16$ of the total energy.
This is regained through work against a small pressure gradient as the flow cools toward the full stop in speed at the loop top; see  \citet{Antokhin04} and \citet{Kee14}.}.

The full ADM model here now specifies the spatial distribution of the X-ray volume emissivity, through the hot component scaling given in equation (\ref{eq:etah}).

\section{Differential Emission Measure}

Let us next consider the X-ray differential emission measure (DEM) resulting from the hot component of this ADM model.
In terms of an emission measure in  electron and proton number density $n_e n_p dV$ within a volume element $dV$, the DEM contribution from a given field line $b$ can be written for a differential segment $db$ along the field line flux tube with area $A$,
\beq
\left [ \frac{d\, EM}{d\, \ln T}  \right ]_b \equiv  n_e n_p \, T \, \frac{dV}{dT} = \frac{T \rho^2 A}{\mu_e \mu_p} \, \frac{db}{dT}
\, .
\label{eq:demdef}
\eeq
Using  the energy equation (\ref{eq:dTdb1}), this can be cast in the form,
\beq
\left [ \frac{d\, EM}{d\, \ln T}  \right ]_b
=  \frac{\rho v A}{\Lambda} \, \frac{5 kT}{2\mubar } 
 ~~ ; ~~ T \le T_s
\, .
\label{eq:demmdot}
\eeq
The qualifier emphasizes that this only applies to temperatures up to the shock temperature $T_s$ for this field line; for $T>T_s$, the DEM is zero. 
Conversely, note that for a given temperature $T$, any field line with $T_s > T$ contributes to the global DEM at that temperature.
This proves very useful for deriving the global DEM in the analysis below.

To proceed, note that, in the ADM model, both $T_s$ and the constant flow tube mass flux $\rho v A$ depend on the footpoint co-latitude of the field line, as set by $\mustar$.
For a differential latitude interval $d\mustar$, the projection of surface radial mass flux along the field implies $\rho v A = d\mustar (d\mdot/d\mustar)$,  which scales with the surface dipole field radial projection $\mu_B$ as \citep{Owocki04c},
\beq
\frac{d\mdot}{d\mustar} = \half \mu_B^2 \Mdot_{B=0} = \frac{2 \mustar^2}{3 \mustar^2 +1 } \, \Mdot_{B=0}
\, ,
\label{eq:dmddmus}
\eeq
where $\Mdot_{B=0}$ is the  standard (CAK) mass loss rate for a  corresponding non-magnetic star, and the factor half accounts for the equal split of the mass loss in the two hemispheres.
Let us then define the {\em cumulative} mass loss in a band $\pm \mustar$ about the equator ($\mustar =0$),
\beqa
\mdot (\mustar )
&\equiv& \int_{-\mustar}^{\mustar} \frac{d\mdot}{d\mustar} \, d\mustar 
\nonumber
\\
&=& \Mdot_{B=0} \int_{0}^{\mustar} \frac{4 \mustar^2}{3 \mustar^2 +1 } d\mustar 
\nonumber
\\
&=& \Mdot_{B=0} \, \frac{4}{3} \,\left [ \mustar - \frac{\arctan (\sqrt{3} \mustar )}{\sqrt{3} } \right ]
\, .
\label{eq:mdotmus}
\eeqa
Note in particular that for a star with a dipole field that is strong enough to retain its dipole form at  the surface for all latitudes, the total surface mass flux is a factor $\mdot(1)/\Mdot_{B=0} = (4/3)(1-\pi/9\sqrt{3})  \approx 0.53$ less than a corresponding non-magnetic, spherical wind. 

For a given cooling parameter $\chi_\infty$ in the ADM model, the shock temperature is a monotonically increasing function of $\mustar$, i.e. $T_s (\mustar , \chi_\infty )$, from which we can derive a corresponding inverse function $\mustar (T_s,\chi_\infty )$.
But for an ADM with a maximum closure radius $R_c$ set by the magnetic confinement parameter $\eta_\ast$, there is a corresponding closure latitude set by $\mu_c = \sqrt{1-\Rstar/R_c}$, with corresponding maximum scaled pre-shock wind speed $w_c = \mu_c^2$. This also sets a maximum post-shock temperature $T_c/T_\infty = w_c^2 = \mu_c^4$, written here in terms of the terminal speed shock temperature $T_\infty$.

In these terms, we can now write the {\em cumulative} fraction of total mass flux that has a shock with temperature above a given threshold $T_s$ as
\beq
p(T_s, \mu_c, \chi_\infty) \equiv \frac{ \mdot (\mu_c ) - \mdot(\mustar(T_s,\chi_\infty))}{\mdot (1) }
\, .
\label{eq:pTsdef}
\eeq
Now, according to (\ref{eq:demmdot}) each field line $b$ contributing to this mass fraction contributes a corresponding amount to the total DEM for {\em any} temperature {\em below} the shock temperature, i.e., for $T \le T_s$. 
This implies that the total DEM from the entire ADM is given in terms of (\ref{eq:demmdot})  by just multiplying by $p(T)$,
\beq
\left [ \frac{d\, {\rm EM}}{d\, \ln T}  \right ]_{tot} (T,\mu_c,\chi_\infty)
= \frac{\mdot (1) \Mdot_{B=0} }{\Lambda} \, \frac{5 kT}{2\mubar }  p(T, \mu_c, \chi_\infty) 
\, .
\label{eq:demtot}
\eeq
Apart from some minor differences in notation and overall scaling, this is essentially equivalent to the result given in equation (8) of \citet{Gayley14}, relating DEM to the shock fraction for the general case of radiatively cooled shocks.
Moreover, for embedded wind shocks arising from the line deshadowing instability of radiative driving, \citet{Cohen14} has recently presented an analysis of X-ray line emission that infers a power-law form for the cumulative distribution $p(T_s)$ for the mass fraction undergoing a shock with temperature $\ge T_s$.

\begin{figure}
\begin{center}
\includegraphics[scale=0.65]{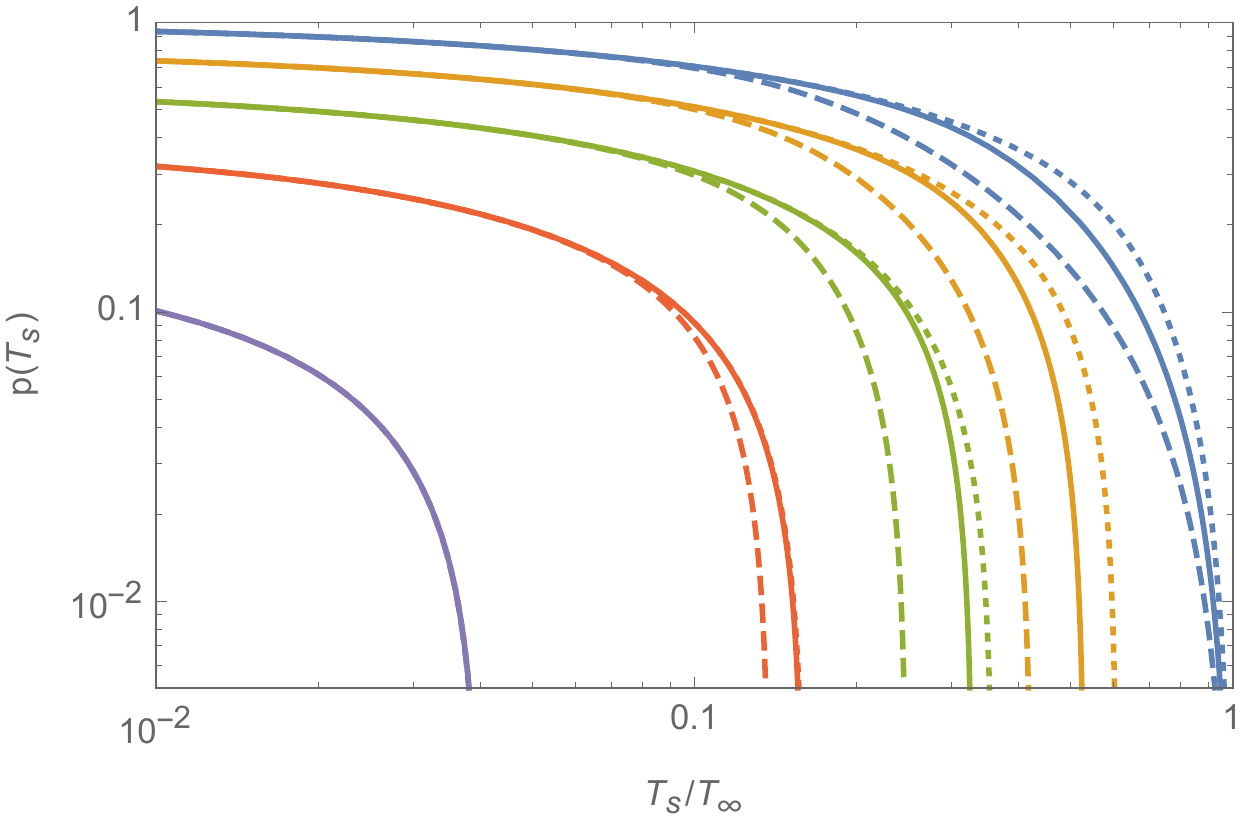}
\includegraphics[scale=0.65]{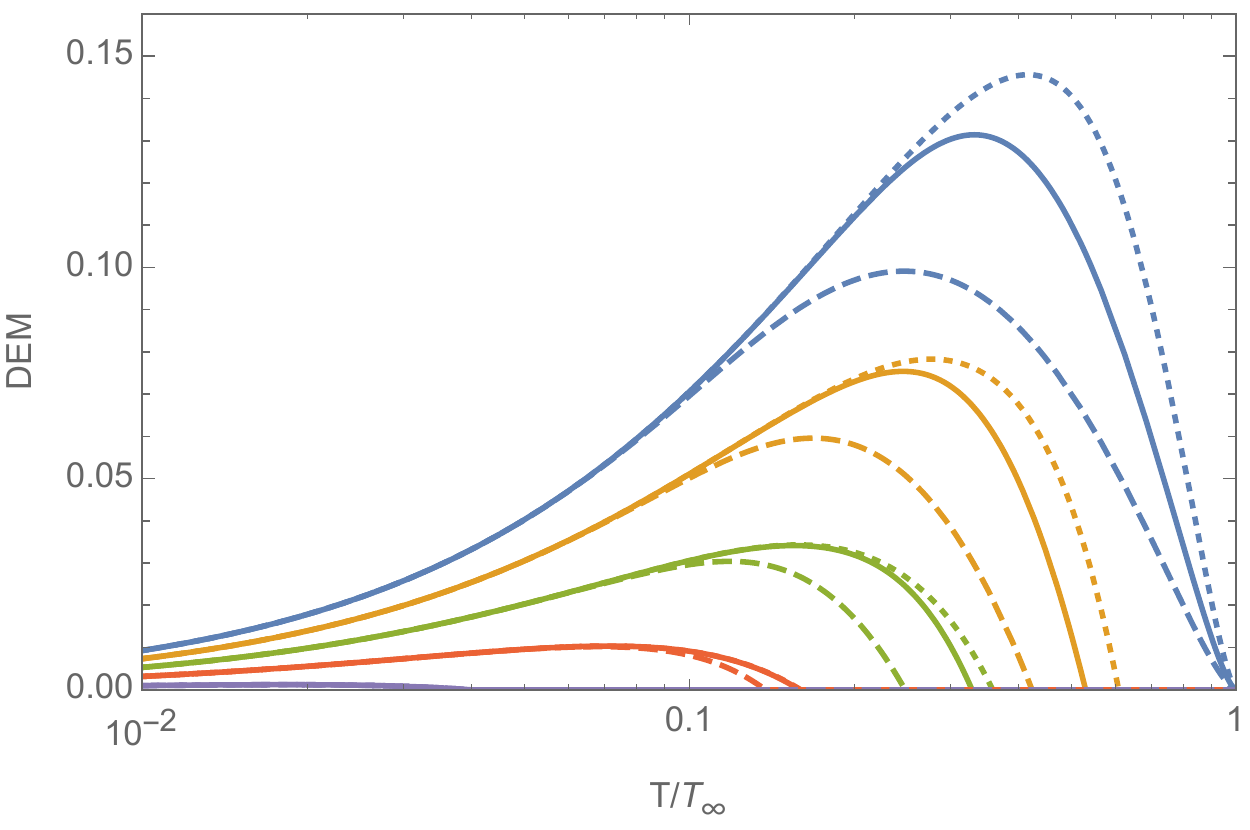}
\caption{
{\em Top:}
Cumulative mass flux fraction $p(T_s)$ yielding a shock temperature  above $T_s$, plotted vs.\ $T_s/T_\infty$ on a log-log scale for scaled closure radius wind speeds $w_c = 1-R/R_c$ from $1$ (uppermost curves) to $0.2$ (lowermost curves) in steps of $0.2$.
The dotted, full, and dashed lines correspond respectively to cooling parameters $\chi_\infty =$ 0.1, 1, and 10.
{\em Bottom:}
Associated scaled differential emission measure, defined  by DEM$\equiv (T/T_\infty) p(T)$, now plotted as linear DEM vs. $\log (T/T_\infty)$, for each of the same parameters cited for the top panel.
}
\label{fig:pTlog}
\end{center}
\end{figure}

The top panel of figure~\ref{fig:pTlog} shows a log-log plot of $p(T_s, \mu_c, \chi_\infty )$ vs. $T_s/T_\infty $ for cooling parameters $\chi_\infty=$ 0.1, 1, and 10 (dotted, full, dashed curves), and closure radii characterized by the maximum scaled wind speed $w_c = 1-\Rstar/R_c$, ranging in steps of -0.2 from 1 (representing the limit $\eta_\ast \rightarrow \infty$ of arbitrarily strong magnetic confinment) to 0.2 (with a near-surface $R_c/\Rstar$ = 1.25, representing only weak magnetic confinement, with $\eta_\ast $ of order unity.)

Note that, for decreasing closure speed $w_c$, the upper temperature cutoff decreases as $T_{max} \sim w_c^2$.
Moreover, for a given, fixed closure speed $w_c$, the temperature cutoff decreases with increasing $\chi_\infty$, 
reflecting the stronger shock retreat from less efficient cooling, giving then slower pre-shock flow speeds, $w_s$, 
and thus a lower maximum shock temperature, $T_{max} \sim w_s^2$.

The bottom panel of figure~\ref{fig:pTlog} shows the corresponding linear-log plots of the scaled differential emission measure. For the simple ADM model here that approximates $\Lambda$ as constant fixed at a value typical of post-shock temperatures, this is defined by 
\beqa
{\rm DEM} (T, \mu_c, \chi_\infty )  &\equiv& \frac{T}{T_\infty} p(T, \mu_c, \chi_\infty )
\nonumber
\\
&=&  \frac{\Lambda}{\mdot (1) \Mdot_{B=0} } \, \frac{2\mubar }{5 kT_\infty}  
\, \left [ \frac{d\, {\rm EM}}{d\, \ln T}  \right ]_{\rm tot} 
 \, .
 \label{eq:demscaled}
\eeqa
This assumed constancy of $\Lambda$ was made to allow analytic solution of the shock retreat from post-shock cooling,  and is justified by the fact that most of the total cooling length occurs from the initial cooling from the post-shock temperature, for which the cooling function varies only weakly with temperature.
But in developing scaling predictions for an observed DEM, one could also readily apply the ADM derived $p(T)$ to a more realistic cooling that includes its full temperature dependence, $\Lambda (T)$, via equation (\ref{eq:demtot}).

\section{H$\alpha$ scaling relation} 
 \label{sec:Halpha} 
 
 \begin{figure}
\begin{center}
\vfill
 \resizebox{\hsize}{!}
	        {\includegraphics[angle=90]{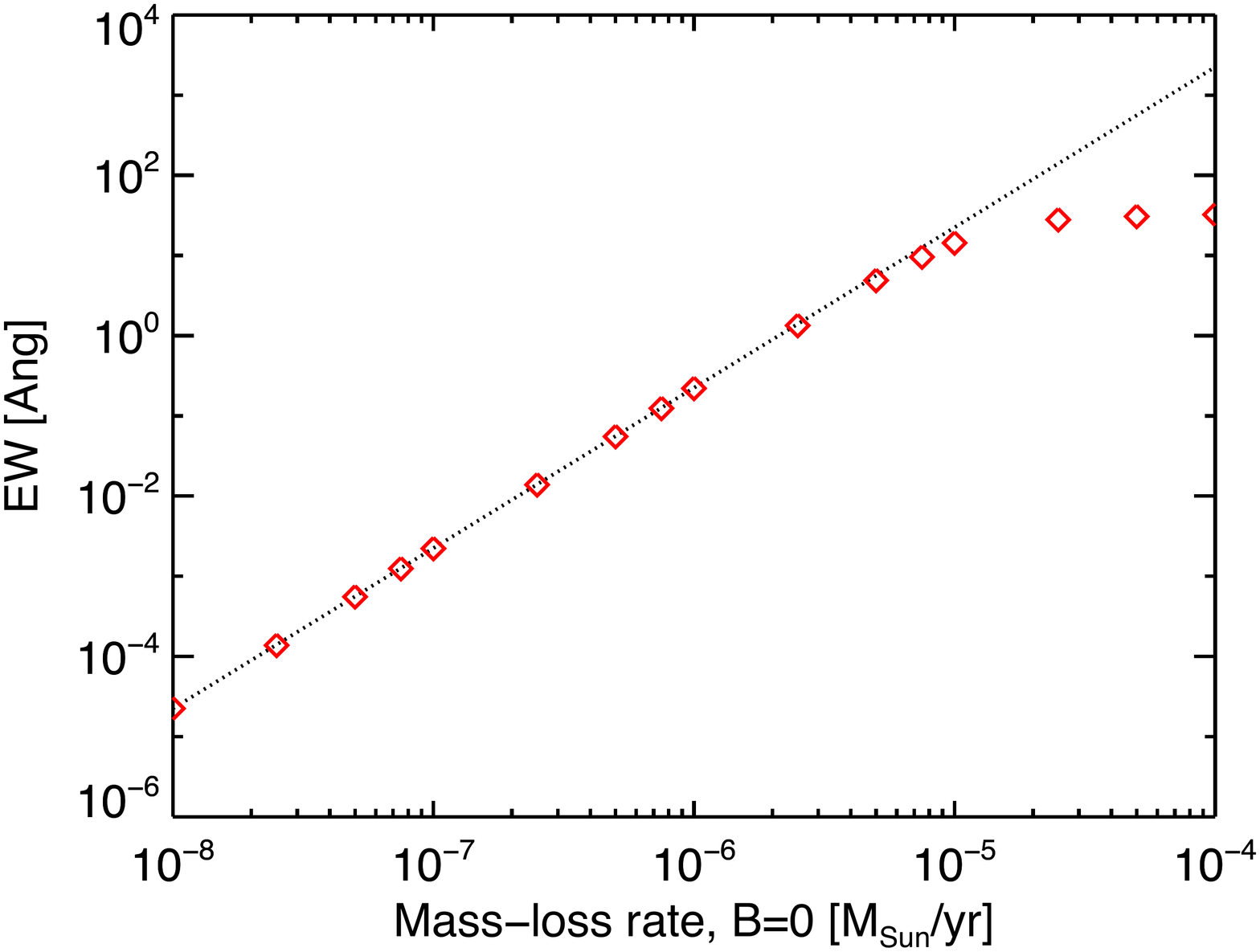}} 
		\centering
\caption{H$\alpha$ emission equivalent width (counted positive) in 
Angstrom vs. $\dot{M}_{\rm B=0}$. The red triangles show results from full radiative transfer calculations using different values of $\dot{M}_{\rm B=0}$ (see text). The dotted line displays the EW scaling according to eqn. C4, scaled to match the result at the lowest mass loss rate. 
}
\label{fig:EW_scaling}
\end{center}
\end{figure}
 
From the Saha-Boltzmann relations and the atomic constants of 
the hydrogen $3 \rightarrow 2$ transition, the parameter $A$ in 
equation (\ref{Eq:tau_sob}) of the main paper can be written as \citep{Puls96, Petrenz96} 
\begin{equation} 
	A = \rm Const.  \ \times \ T_{\rm e}^{-3/2} \frac{1+Y_{\rm He} I_{\rm He}}{(1+ 4Y_{\rm He})^2} 
	\left ( b_{\rm 2} e^{3.95/T_{\rm e}} - b_{\rm 3} e^{1.753/T_{\rm e}} \right ).     
\end{equation} 
In this equation $T_{\rm e}$ is the wind electron temperature, $Y_{\rm He} \equiv n_{\rm h}/n_{\rm He}$ the  helium number fraction with respect to hydrogen, and  $I_{\rm He}$ the number of free electrons per helium ion. 
$b_{\rm i} \equiv n_{\rm i}/n_{\rm i}^*$ further denotes the kinetic equilibrium (`NLTE') departure coefficient of atomic level $i$, for population 
number density $n_{\rm i}$ and LTE density  $n_{\rm i}^*$ with respect to the \textit{real} population of the ground state of the next higher ion \citep[see][]{Mihalas78}. 

To derive a characteristic scaling relation for the strength of H$\alpha$ emission, let us for now neglect the influence of the photospheric absorption profile and assume an LTE source function set by the radiation temperature of the star,  so that  the absorption and emission in front of the stellar disc cancel. 
Using further the fact that the H$\alpha$ transition is optically thin in most parts of the wind for O-stars \citep{Puls96}, we can write down a `Sobolev emission measure' scaling for a clumping-corrected ADM model in terms of integrals over impact parameter $p$ and normalized 
frequency $x = (\nu/\nu_0 -1 )c/v_{\rm e}$, 
\begin{equation} 
   W_{\rm ADM} \propto
\int_{0}^{\infty} p \, dp \, \int_{0}^{1} dx \,  
\left[  \rho^2 f_{\rm cl} v_{\rm e}/(dv_{\rm n}/dn)  \right ]_{x=v_{\rm n}/v_{\rm e}} ,
  \label{Eq:em_meas}
\end{equation}
wherein the density $\rho$ and the projected 
velocity gradient $dv_{\rm n}/dn$ along line of sight direction ${\hat n}$
are to be evaluated at the resonance location $x =
v_{\rm n}/v_{\rm e}$ for each frequency $x$.

For an observer viewing from a direction ${\hat n}={\hat z}$ along the magnetic pole, the resonance condition $x =
v_{\rm z}/v_{\rm e}$ will then be close to the magnetic equator, with $p=r$ and
 $dv_{\rm z}/dz \approx v_{\rm e}/\hzd$. 
From (\ref{eq:rhoc2}), the cooled downflow density in this region is
\beq
\rho_c (r,\mu=0)  = \frac{\dot{M}_{\rm B=0}}{4 \pi R_\ast^2 v_{\rm e}}
 \frac{ \sqrt{ r/\Rstar - 1} }
{\hzd/r \, (4r/\Rstar -3)} \,
\left ( \frac{\Rstar}{r} \right )^{2} \, .
\label{eq:rhoeq}
\eeq
Applying this and the velocity gradient scaling in equation (\ref{Eq:em_meas}) gives for the principal scaling of emission measure,
\begin{equation} 
  W_{\rm ADM} \propto \frac{\dot{M}_{B=0}^2} {\Rstar^3 v^2_{\rm e}} 
   \frac{f(R_c)}{\hzd/R_\ast}.
  \label{Eq:Q}
\end{equation}
where the last part describes the dependence on the size of the magnetosphere in terms of the function 
\begin{equation} 
	f(R_c) = \int_1^{R^{'}_{c}} \frac{r'-1}{r'(4r'-3)^2} dr', 
	\label{Eq:adm_ew}
\end{equation} 
where $r' \equiv r/R_\ast$ and $R'_c \equiv R_c/R_\star$.
Equation \ref{Eq:Q}  thus suggests the standard H$\alpha$ emission scaling  for non-magnetic OB-stars \citep[e.g.,][]{Puls08} is modified in the ADM model by the smoothing length $\hzd$ and a function $f(R_c)$ describing the size of the ADM; these parameters then account for the influence of the magnetic field on the wind structure and emission measure.

While the formation process of H$\alpha$ and other optical emission lines in magnetic OB-star winds will in reality be much more complex than discussed here, comparison to full radiative transfer computations using the ADM flow structure and 3D formal solver described in \S\ref{sec:firstdiag}
nonetheless captures quite well the principal scalings of the emission. 
Fig. C1 illustrates this, comparing H$\alpha$ emission equivalent widths from such 
full computations with the simple scaling relation eqn. C4.
All models here have been calculated for a polar observer using 
the same basic set-up as in \S4.2; to allow for a simple comparison with 
the predicted scaling relation, however, we now neglect the photospheric 
absorption profile, assume an LTE source function,  
and  only vary $\dot{M}_{\rm B=0}$ while keeping 
all other input parameters fixed (at same values as in \S4.2, including $R_{\rm C} = 3.5 R_\star$).

The figure shows the polar-vew optically thin scaling is followed perfectly 
for typical OB-star mass-loss rates, but breaks down in the optically thick region of 
higher $\dot{M_{\rm B=0}}$. In this regime the emission 
instead simply scales with the projected surface area of the ADM.
The $\dot{M}_{\rm B=0}$ derived for HD191612 in \S4.2 is close to where the scaling in Fig. C1 begins to fall, so
further studies are needed to determine more precisely   
under which conditions and phases such scaling might apply.  
Also, recall here that eqn. C4 is derived for an observer viewing from 
above the magnetic pole, and so only provides scaling-information regarding 
the \textit{level} of H$\alpha$ line-emission during near-polar phases; the scaling 
is analogous to how line-emission is used to derive empirical mass-loss rates 
in non-magnetic stars, and does not provide any predictions for the 
\textit{rotational phase variation} of the emission (which rather is set primarily 
by the magnetic geometry; see also discussion in main text).

\end{document}